\documentclass[a4paper,fleqn,usenatbib]{mnras}

\usepackage[T1]{fontenc}
\usepackage{ae,aecompl}

\usepackage{graphicx}	
\usepackage{amsmath}	
\usepackage{amssymb}	
\usepackage{mathtools}

\usepackage{cleveref}

\let\mr=\mathrm

\newcommand{\bq}{\begin{equation}}
\newcommand{\eq}{\end{equation}}

\newcommand{\dd}{\, \mathrm{d}}

\newcommand{\LCDM}{\Lambda\mathrm{CDM}}

\newcommand{\hpM}{h\mathrm{Mpc}^{-1}}
\newcommand{\Mpc}{\mathrm{Mpc}^{-1}}

\def\equationautorefname~#1\null{equation~(#1)\null}
\def\sectionautorefname~#1\null{Section~#1\null}
\def\figureautorefname~#1\null{Fig.~#1\null}

\title[Euclid: inflation features]{Constraints on features in
the inflationary potential from future Euclid data}

\author[I. Debono et al.]{Ivan Debono,$^{1}$\thanks{E-mail: mail@ivandebono.eu}
Dhiraj Kumar Hazra,$^{2,3,4}$\thanks{E-mail: dhiraj@imsc.res.in}
Arman Shafieloo,$^{5,6}$ \newauthor
George F. Smoot,$^{1,7,8}$ 
Alexei A. Starobinsky$^{9}$
\\
$^{1}${Paris Centre for Cosmological Physics, Universit\'e de Paris,
CNRS, Astroparticule et Cosmologie, F-75006 Paris, France}\\
$^{3}${The Institute of Mathematical Sciences, HBNI, CIT Campus, Chennai 600113, India}\\
$^{3}${Osservatorio di Astrofisica e Scienza dello Spazio di Bologna/Istituto Nazionale di Astrofisica, via Gobetti 101, I-40129 Bologna, Italy}\\
$^{4}${Istituto Nazionale Di Fisica Nucleare, Sezione di Bologna,Viale Berti Pichat, 6/2, I-40127 Bologna, Italy}\\
$^{5}$Korea Astronomy and Space Science Institute, Daejeon 34055, South Korea \\
$^{6}$Astronomy \& Space Science Department, Korea University of Science and Technology, Daejeon 34113, South Korea\\
$^{7}${Physics Department and Lawrence Berkeley National Laboratory, University of California, Berkeley, 94720 CA, USA}\\
$^{8}${Institute for Advanced Study, Hong Kong University of Science and Technology, Clear Water Bay, Kowloon, 999077, Hong Kong}\\
$^{9}${Landau Institute for Theoretical Physics RAS, Moscow, 119334, Russia }}

\date{Accepted 2020 June 16}

\pubyear{2020}

\begin{document}
\label{firstpage}
\pagerange{\pageref{firstpage}--\pageref{lastpage}}
\maketitle

\begin{abstract}
With \textit{Planck} cosmic microwave background observations, we established the spectral amplitude and tilt of the primordial power spectrum. Evidence of a red spectral tilt ($n_\mathrm{s}=0.96$) at $8\sigma$ provides strong support for the inflationary mechanism, especially the slow-roll of the effective scalar field in its nearly flat potential as the generator of scalar primordial perturbations. With the next generation of large-scale structure surveys, we expect to probe primordial physics beyond the overall shape and amplitude of the main, smooth and slowly-changing part of the inflaton potential. Using the specifications for the upcoming \textit{Euclid} survey, we investigate to what extent we can constrain the inflation potential beyond its established slow-roll behaviour. We provide robust forecasts with \textit{Euclid} and \textit{Planck} mock data from nine fiducial power spectra that contain suppression and wiggles at different cosmological scales, using the Wiggly Whipped Inflation (WWI) framework to generate these features in the primordial spectrum. We include both \textit{Euclid} cosmic shear and galaxy clustering, with a conservative cut-off for non-linear scales. Using Markov chain Monte Carlo simulations, we obtain an improvement in constraints in the WWI potential, as well an improvement for the background cosmology parameters. We find that apart from improving the constraints on the overall scale of the inflationary potential by 40-50 per cent, we can also identify oscillations in the primordial spectrum that are present within intermediate to small scales ($k\sim0.01-0.2\,\mathrm{Mpc^{-1}}$). 
\end{abstract}




\section{Introduction}

Cosmology is at a point in its history where observations have caught up with theories, and physics at the largest cosmological scales is probed in full-sky surveys. Data from various observations allow us to measure the parameters in our cosmological model with increasing precision. These data include cosmic microwave background (CMB) measurements such as \textit{WMAP} \citep{WMAP9} and \textit{Planck} \citep{Planck-Collaboration:2013aa}, supernovae compilations (e.g. \citealt{Goldhaber:2009aa}, SCP), large-scale structure maps (e.g. \citealt{Ahn2014}, SDSS), and weak-lensing observations (e.g. \citealt{Parker2007}; \citealt{Schrabback:2009}). The next generation of observations, such as \textit{Euclid} (\citealt{laureijs2011euclid}; \citealt{Amendola2018}), the Square Kilometre Array (SKS, \citealt{Blake2004}; \citealt{maartens2015ska}; \citealt{santos2015ska}), and the Large Synoptic Survey Telescope (LSST, \citealt{LSST2019})  are expected to provide order of magnitude improvements in precision, and in the ability to constrain different cosmological processes. 

The fundamental questions facing modern-day cosmologists are not simply about parameter estimation in a known model, but about the possibility of new physics. They are questions about model selection. In addition to estimating the values of the parameters in the model, this involves decisions on which parameters to include or exclude. In some cases, the inclusion of parameters is possible only by invoking new physical models.

The $\Lambda$ cold dark matter ($\LCDM$) concordance model can fit different astrophysical datasets with only six parameters describing the mass--energy content of the Universe (baryons, CDM and a cosmological constant or constant dark energy) and the initial conditions. Any deviations from $\LCDM$ are too small compared to the current observational uncertainties to be inferred from cosmological data alone. However, it does not mean that additional parameters are ruled out. 

There are several open questions in modern cosmology. Most of the matter in the Universe is dark matter, whose nature is not known. Another open question is the nature of the component causing the accelerated expansion of the Universe. The data are compatible with a cosmological constant, but do not exclude dynamical dark energy. Finally, there is the question of the conditions in the very early Universe. 

In this paper we focus on the physics of the primordial Universe, and examine the ability of \textit{Euclid}  to provide information about features in the primordial power spectrum beyond that which \textit{Planck} has provided. In \citet{2012JCAP...04..005H} and \citet{2016JCAP...10..041B}, it was found that for models with features, a large-scale structure survey like \textit{Euclid} will be essential to detect and measure these features.  With CMB probes, we measure the angular power spectrum of the anisotropies in the two-dimensional multipole space. This is a projection of the power spectrum in the three-dimensional momentum space. This projection smooths out narrow features in momentum-space at large multipoles. This does not occurs with large-scale structure surveys like \textit{Euclid}

This work is motivated by the above consideration. \textit{Planck} and \textit{Euclid} show substantial overlap in their ability to probe cosmological scales. With \textit{Planck}, we have not been able to find strong evidence for the existence of features in the primordial power spectrum. Certain types of features in the intermediate and small scales persist in all data releases in \textit{Planck}. Their existence, if detected, will directly identify the fine shape of the inflaton potential, its transitions, and its nature. 

In the last four decades, various types of features in the primordial power spectrum generated by local or non-local modification in the potential have been proposed (e.g. \citealt{Starobinsky:1992ts}; \citealt{1996ApJ...464L...1B}; \citealt{2003ApJS..148..213P}; \citealt{GARIAZZO201738}; \citealt{PhysRevResearch.1.033209}; see also \citealt{ListOfInflationModels}). In this work, we aim to forecast parameter constraints for different types of primordial features appearing at different cosmological scales. Should the inflation potential really result in a primordial power spectrum with features, to what extent can \textit{Euclid} observations probe these features? 

In order to explore the features we use Wiggly Whipped Inflation (WWI; \citealt{Hazra2014}), which can generate a wide variety of primordial power spectra with features at different cosmological scales which otherwise can be obtained using different potentials. Due to its generic nature, the WWI framework, being confronted with \textit{Planck} temperature and polarization data, was capable of offering a family of primordial power spectra that provided better fit to the combined data compared to the nearly scale-invariant spectrum \citep{Hazra2016}. By using Wiggly Whipped Inflation, we allow for a broad range of primordial power spectrum features in a single framework. So far, three types of features are known to provide improvement of fit to the data compared to power law primordial spectrum, namely, large-scale suppression or dip, intermediate scale oscillations near first acoustic peak and near $\ell=600-800$ and certain high-frequency oscillations that continue towards small scales. 

In this paper, we use the best-fitting Wiggly Whipped Inflation models obtained using \textit{Planck} to create fiducial cosmologies and data for \textit{Planck} and \textit{Euclid}. We use Markov chain Monte Carlo (MCMC) methods to forecast the ability of \textit{Euclid} observations to add information and the possibility of identifying different features in the primordial power spectrum.
 
This paper is organized as follows. In \autoref{sec:PrimordialPhysics}, we describe the different models for the primordial power spectrum considered in our work, and provide details of the Wiggly Whipped Inflation potential. We also explain how WWI is supported by the data. In \autoref{sec:Method}, we describe the methods we use to generate mock data for \textit{Planck} CMB, \textit{Euclid} cosmic shearand \textit{Euclid} galaxy clustering, together the survey specifications and the theoretical error modelling. Our cosmological models and the various software codes used in this work are described in this section. We present our results in \autoref{sec:Results}. Finally, we provide some concluding remarks and perspectives for future work in \autoref{sec:Conclusions}.

\section{Primordial physics}
\label{sec:PrimordialPhysics}

The large-scale structure that we observe today in the Universe is seeded by primordial quantum perturbations. These quantum fluctuations originated and evolved during the inflationary epoch. The form of the primordial power spectrum describing these perturbations depends on the inflation potential.
Here we focus on the physics of the primordial Universe, and examine the ability of \textit{Euclid} to provide information about features in the primordial power spectrum.  

\subsection{Power law primordial spectrum in the Concordance Model}

The simplest primordial power spectrum, which is the one used in the six-parameter Concordance Model of cosmology, is a power law with the following phenomenological form:
 \bq 
 P_\mathrm{S}^{\mathrm{Plaw}}(k)=A_\mathrm{s}\left(\frac{k}{k_{0}}\right)^{n_{\mathrm{s}}-1} \ ,
 \eq 
 where $A_\mathrm{s}$ is the amplitude and $n_{\mathrm{s}}$ is the tilt
 of the spectrum of primordial perturbations (see e.g.~\citealt{Kosowsky:1995, Bridle:2003}). In a scale-invariant power spectrum, $n_{\mathrm{s}}=1$. However, by now this spectrum is firmly excluded by observations.
  
Based on this model, \textit{Planck}-2018 analysis reports $n_{\mathrm{s}}=0.9649\pm0.0042$. This $8\sigma$ statistical difference between the data and the scale-independent primordial power spectrum model is among the most remarkable results of the \textit{Planck} mission~\citep{Planck-Collaboration:2013aa,Planck-Collaboration2015aa,Planck2018:params,Planck2018:inflation}. 

The scale-invariant power spectrum is featureless. Broad features in the power spectrum can be described by logarithmic derivatives of the tilt (running and running-of-running), or by local and non-local wiggles in the power spectrum. Any variation beyond the tilt in the spectrum has not been established to date with any statistical significance. 
 
However, since the initial release of \textit{WMAP}~\citep{WMAP:2003}, through to different releases of \textit{Planck} data, the $\LCDM$ model has shown certain outliers. In different model-dependent and independent reconstructions~\citep{Hannestad_2001,Tegmark_2002,Bridle:2003,Mukherjee_2003,Shafieloo_2004,Kogo_2005,Leach:2005av,TocchiniValentini:2005ja,Shafieloo_2008,Nicholson_2009,Paykari_2010,Gauthier_2012,Hlozek_2012,V_zquez_2012,Hazra_2013,Dorn_2014,Hazra_2014,Hazra2014bJCAP,Hunt_2014} primordial features at particular scales have been found to address these outliers. While the statistical significance of these outliers is rather low ($3\sigma$), they interestingly persisted in all releases of these two full-sky surveys and at the same angular scales \citep{Hazra2014bJCAP}. If we do not attribute the occurrence of these outliers to the Concordance Model as statistical, the low significance is then expected to come from the insufficient signal-to-noise ratio. With 3D surveys such as \textit{Euclid}, we look forward to joint estimation with CMB data where the existence of any outliers can be explained by detectable features in the spectrum. Projected constraints on broad features that are described by running and running-of-running of spectral index in the context of \textit{Euclid} were studied in~\citet{Debono:2009} using Fisher analysis. In this work, we use the MCMC method to forecast the constraints on possible oscillations in the primordial spectrum. Instead of just imposing some parametric modification in the spectrum we model the existence of such features from inflation theory and therefore we project constraints in the shape of the inflationary potential.

\subsection{Wiggly Whipped Inflation}

The aforementioned form of power-law primordial power spectrum is a prediction of inflation where the scalar field (inflaton) slowly rolls down to the bottom of the flat inflationary potential. With the constraints on the tilt and an upper bound on the amplitude of tensor perturbation with respect to scalar perturbation, various surveys have ruled out a wide class of models. However, fundamental questions such as the energy scale of inflation and the detailed shape of the potential remain unanswered. Note that any changes in the nearly flat potential will eventually lead to certain features in the spectrum. 
Local glitches in the potential including rapid change of its amplitude,
or the break in its first or second derivatives \citep{Starobinsky:1992ts,Starobinsky1998,StepFeatures3,
StepFeatures4,Joy:2007na,Joy:2008qd,StepFeatures1,StepFeatures5,StepFeatures6,StepFeatures2,StepFeatures7}, false vacuum decay (leading to open inflation, in
particular) \citep{Linde:1998iw,Linde:1999wv,Bousso:2014jca}, or an
inflection point in the potential \citep{Allahverdi:2006wt,Jain:2008dw}, or oscillations in the potential~\citep{Ashoorioon:2006wc,Biswas:2010si,Flauger:2009ab,Pahud:2008ae,Aich:2011qv,Hazra:2012vs,Peiris:2013opa,Meerburg:2013dla,Easther:2013kla,Motohashi:2015hpa,Miranda:2015cea} all lead to local and non-local oscillations in the spectrum. Direct reconstruction of the primordial spectrum from the \textit{Planck} data~\citet{Hazra2014bJCAP} hints at large-scale oscillations, an intermediate-scale burst of oscillations, and persistent high-frequency oscillations within intermediate to small scales. While these types of features can be obtained by different potentials, in this work we will be using the Wiggly Whipped Inflation, which is known the provide these local and non-local features in a unified framework.

Wiggly Whipped Inflation was first proposed in~\citet{Hazra2014} as an extension of the Whipped Inflation model introduced in~\citet{Hazra2014b}. Both belong to the class of models with a large field inflaton potential. In Whipped Inflation, the inflaton starts with a power-law potential. After an initial period of relatively fast roll that lasts until after a few e-folds inside the horizon, it transits to the attractor of the slow-roll part of the potential with a lower power. The initial motivation for Wiggly Whipped Inflation was the \textit{BICEP-2} result announced by the~\citet{BICEP2-Collaboration:2014} and~\citet{Ade:2014} where the suppression of scalar power at large scales with appropriate tensor power spectrum amplitude ruled out the simplest power-law form of the spectrum in the \textit{Planck}-\textit{BICEP2} joint analysis. The data were subsequently re-interpreted in the~\citet{BICEP2_Keck:2015}, where dust polarization amplitude could consistently describe the observed B modes at large scales. This re-interpretation reduced the statistical significance of WWI associated with large field models. Thereafter, in~\citet{Hazra2016} the authors redesigned the potential in the light of new data. With \textit{Planck} temperature and polarization data, using only two potentials in the WWI framework, the authors identified five types of spectra that provided improvement in fit to the \textit{Planck} data compared to power-law spectrum by a $\Delta\chi_\mathrm{eff}=$ 12--14 with 2--4 extra parameters.

\subsubsection{The inflationary potential}
\label{sec:InflationaryPotential}

In the WWI framework, the two potentials we consider are Wiggly Whipped Inflation (WWI potential hereafter) and Wiggly Whipped Inflation Prime (WWIP potential hereafter).

The WWI potential is defined by the equation:
\begin{equation}
V({\phi})=V_{i} \left(1-\left(\frac{\phi}{\mu}\right)^{p}\right)+\Theta(\phi_{\mr T}-\phi) V_{i}\left(\gamma (\phi_{\mr T}-\phi)^{q}+\phi_{0}^q\right),~\label{eq:equation-WWI}
\end{equation}
where we note that $V_{S}(\phi)=V_{i} \left(1-\left(\frac{\phi}{\mu}\right)^{p}\right)$ has two parameters, $V_{i}$ and $\mu$. The parameter $\mu$ and the index $p$ determine the spectral tilt $n_{\mr s}$ and the tensor-to-scalar ratio $r$.

We choose the values $p=4$ and $\mu=15~M_\mathrm{P}$, where $M_\mathrm{P}=1$ is the reduced Planck mass, such that $n_{\mr s}\sim0.96$ and $r\sim{\cal O}(10^{-2})$ (as in~\citealt{Efstathiou:2006ak}).
The transition and discontinuity happen at the field value $\phi_{\mr T}$.
In this case, a featureless primordial power spectrum is obtained if $\gamma=0$ and $\phi_{0}=0$. The Heaviside Theta function $\Theta(\phi_{\mr T}-\phi)$ is modelled numerically as usual by a Tanh step
($\frac{1}{2}\left[1+{\tanh}[{(\phi-\phi_{\mr T})}/{\delta}]\right]$) and thereby introduces a new extra parameter $\delta$.

The WWIP potential is defined by:
\begin{align}
V({\phi})=\, & \Theta(\phi_{\mr T}-\phi) V_{i}
\left(1-\exp\left[-\alpha\kappa\phi\right]\right)\nonumber \\
&+\Theta(\phi-\phi_{\mr T})
V_{ii}\left(1-\exp\left[-\alpha\kappa(\phi-\phi_{0})\right]\right)\, . ~\label{eq:equation-WWIP}
\end{align}
This potential is same as that used in \citet{Hazra2016}. It is composed of $\alpha$-attractor potentials \citep{alpha_attractor}, which include the Einstein frame effective potential of the Starobinsky
$R+R^2$ inflationary model \citep{STAROBINSKY198099} as a particular case for $\alpha=\sqrt{2/3}\approx 0.816$, where
$R$ is the Ricci scalar, with different slopes appearing in the exponent, allowing a discontinuity in the derivative. Since in this case the potential is continuous, $V_{ii}$ can be derived from $V_{i}$. In the WWIP models considered in this paper, we set $\alpha=\sqrt{2/3}$. In our convention, $\kappa^2=8\pi G$ is equal to $1$. The parameter $G$ is the gravitational constant.

\section{Method}
\label{sec:Method}

Our forecasts use the MCMC technique, with mock data from fiducial cosmological models. We begin by giving brief details of the simulated \textit{Euclid} and \textit{Planck} data sets used in our forecasts. The \textit{Euclid} likelihoods used in this paper are described exhaustively in \citet{Sprenger2018}.

\subsection{The simulated data}
\label{sec:Data}

Data from the \textit{Euclid} mission are not yet available, so we compute mock data from a fiducial cosmology following the method defined in~\citet{Sprenger2018}. Since our aim is to quantify the improvement in constraints from future \textit{Euclid} data, we carry out two MCMC forecasts for each cosmological model: first with simulated \textit{Planck} CMB data alone, then with joint \textit{Euclid} galaxy clustering and cosmic shear, and \textit{Planck} CMB data. 

\subsubsection{Cosmic microwave background}
\label{sec:CMB}

For \textit{Planck}, we run our forecasts with mock temperature, polarization, and CMB lensing data generated for the parameter values of the fiducial model. We use the fake realistic \textit{Planck} likelihood provided with the publicly-available \textsc{montepython} package, which models the full \textit{Planck} mission. It is based on the fake \textit{Planck} Blue Book likelihood, which was modelled on  the 2005 Blue Book \citep{collaboration2006scientific}. We use a multipole range $2\le\ell \le 3000$, with $f_\text{sky}=0.57$. We do not include B modes. 

This likelihood uses noise spectra from \citet{BrickmannHooper2019}, which match the sensitivity of the final \textit{Planck} data release. We verified this by running MCMC simulations using \textsc{montepython} with \textsc{class} for real \textit{Planck} 2015 data. We repeated the same simulations using \textsc{cosmomc} \citep{CosmoMC1,CosmoMC2} with \textsc{camb} \citep{CAMB2000,CAMB2012}. In both cases, we found excellent agreement between the results from the simulated data and the real data. The repetition of the process with two software packages using different codes for the calculation of the power spectra validated the numerical accuracy of the codes.

\subsubsection{Cosmic shear}
\label{sec:CS}

In this section we give a brief description of the relevant quantities in our cosmic shear likelihood. Further details are found in 
\citet{Sprenger2018}, and references therein. This code, and the code for the galaxy clustering likelihood are both publicly-available in the \textsc{montepython} package.

Cosmic shear surveys map the alignments in the distortion of galaxies caused by weak gravitational lensing as a result of density inhomogeneities along the line of sight. It provides an effective way to map dark matter, and is therefore a powerful probe of large-scale structure. Cosmological information is extracted from auto-correlations and cross-correlations of alignment maps at different redshifts (see e.g. \citealt{2010CQGra..27w3001B})

The matter power spectrum is defined as:
\bq 
 \langle \delta(\mathbfit{k})\delta^\star(\mathbfit{k}')\rangle= {(2\pi)}^3\delta_D^3(\mathbfit{k}-\mathbfit{k}') P(k) \ .
 \eq 
 
The three-dimensional matter power spectrum is projected onto a two-dimensional lensing correlation function for redshift bins $i$ and $j$ at multipoles $\ell$:
\bq
\label{Cl_Pk}
C_\ell^{ij} = \frac{9}{16}\Omega_m^2H_0^4 \int_0^{\infty}\frac{\dd r}{r^2} g_i(r) g_j(r) P\left(k=\frac{\ell}{r},z(r)\right) \ .
\eq
The functions $g_i(r)$ depend on the radial distribution of galaxies in the redshift bin $i$.

A noise term  $N_{\ell}$ is added to the theoretical $C_\ell^{ij}$ due to the intrinsic alignment of galaxies. The noise spectrum is:
\bq
N_{\ell}^{ij} = \delta_{ij}\sigma_{\text{shear}}^2 n_i^{-1} \ ,
\eq
where $\sigma_{\text{shear}}$ is the root mean square of the galaxy intrinsic ellipticity. We set this to $0.3$.
The term $n_i$ is the number of galaxies per steradian in redshift bin $i$. We divide the redshift range into 10 redshift bins, with an equal number of galaxies in each. Therefore, for every redshift bin we have:
\bq
n_i = \frac{n_{\text{gal}}}{10} \times 3600\left(\frac{180}{\pi}\right)^2 \ ,
\eq
where the number of observed galaxies $n_{\text{gal}}=30\,\text{arcmin}^{-2}$. 

\subsubsection{Cosmic shear likelihood}
\label{sec:CS_likelihood}

To calculate the cosmic shear likelihood, we use the method described in \citet{Sprenger2018}, which in turn is taken from \citet{Audren2013}. This method defines the likelihood as:
\bq
\label{LensingLikelihood}
-2\ln{\cal L} \equiv \sum_\ell(2\ell+1)f_{\text{sky}}\left(\frac{d_\ell^{\text{mix}}}{d_\ell^{\text{th}}}+\ln\frac{d_\ell^{\text{th}}}{d_\ell^{\text{obs}}}-N\right) \ ,
\eq
where `obs' and `th' denote observed and theoretical quantities, respectively.
The term $N$ is the number of redshift bins.  Each $C_\ell$ matrix has dimension $N$, and the matrix determinants are denoted by $d$. We have three kinds of determinant: the determinant of the theoretical angular power:
\bq
d_\ell^{\mr{th}} = \det \left( C_\ell^{{\mr{th}}\,ij} + N_\ell^{ij} \right) \ ,
\eq 
that of the observed angular power spectrum:
\bq
d_\ell^{\mr{obs}} = \det \left( C_\ell^{{\mr{fiducial}}\,ij} + N_\ell^{ij} \right) \ ,
\eq 
and a mixed determinant:
\bq
d_\ell^{\mr{mix}} = \sum_k \det\left( N_\ell^{ij} + \begin{cases} C_\ell^{{\mr{th}}\,ij} &, j\neq k\\[10pt] C_\ell^{{\mr{fiducial}}\,ij} &, j=k \end{cases} \right)
\eq
Note that the observed and theoretical spectra include a noise term $N_\ell^{ij}$. 

In our MCMC simulations, the sampled points in parameter space act as our observed power spectra, while the theoretical power spectrum is produced using the fiducial model.

\subsubsection{\textit{Euclid} cosmic shear specifications}
\label{sec:EuclidCS_specs}

We use the number density of galaxies with the corresponding redshift errors taken from \citet{Audren2013} and used in \citet{Sprenger2018}, where the unnormalized redshift number density distribution is defined by:
\bq
\frac{\dd n_{\text{gal}}}{\dd z} = z^{\beta} \exp\left[-\left(\frac{z}{\alpha z_\mathrm{m}}\right)^{\gamma}\right] \ .
\eq 
We set the values $\alpha=\sqrt{2}$, $\beta=2$, and $\gamma=1.5$.
In this equation, $z$ is the redshift, while $z_\mathrm{m}=0.9$ is the median redshift of the sources.

The redshift uncertainty is parametrized by a Gaussian error which depends on the redshift $z$. For redshifts up to $z_{\text{photomax}}$, the redshift uncertainty is a function of the photometric redshift error $\sigma_{\text{photo-z}}=0.05$. Beyond $z_{\text{photomax}}$, we assign a larger error $\sigma_{\text{no-z}}=0.3$. We use a value of $z_{\text{photomax}}=4$. The details of the redshift uncertainty parametrization are found in  \citet{Harrison2016} and \citet{Sprenger2018}.

The sky coverage for \textit{Euclid} $f_{\text{sky}}= 0.3636$, and we use the same value for galaxy clustering.

\subsubsection{Galaxy clustering}
\label{sec:GC}

Galaxies are not randomly distributed in space, but tend to be found in clusters. 
The galaxy power spectrum is defined as a function of a continuous density field, which represents
the probability density $p_\mathrm{g}$ of finding a galaxy at some position $\mathbfit{r}$. The galaxy density perturbation $\delta_\mathrm{g}$ is therefore a perturbation of this probability density:
\bq
p_\mathrm{g}(\mathbfit{r}) = \bar{n}(\mathbfit{r})(1+\delta_\mathrm{g}(\mathbfit{r})) \ .
\eq
The quantity $\bar{n}(\mathbfit{r})$ is the expected number density of galaxies on a homogeneous background, calculated as the mean density over a sufficiently large volume. In our galaxy clustering calculation, this will be the volume corresponding to one redshift bin.
The spatial distribution of galaxies represents a biased tracer of the underlying dark matter distribution, so the conversion from the matter to the galaxy power spectrum must take into account various effects. We use the method developed in \citet{Sprenger2018}.

The observed galaxy power spectrum $P_\mathrm{g}$ is related to the matter power spectrum $P_\mathrm{m}$ by:
\begin{align}
\label{eq:P_g-def}
P_\mathrm{g}(k,\mu,z) =&  f_{\text{AP}}(z) \times f_{\text{res}}(k,\mu,z) \\
& \times f_{\text{RSD}}(\hat{k},\hat{\mu},z) \times b^2(z) \times P_\mathrm{m}(\hat{k},z) \nonumber \ .
\end{align}
This relation assumes a flat-sky approximation \citep{Lemos2017,Asgari2018}, which allows us to define the angle between the Fourier modes
 $\mathbfit{k}$ and the line-of-sight distance vector $\mathbfit{r}$. Thus, in \autoref{eq:P_g-def}, 
 \bq
k = \vert\mathbfit{k}\vert \ ,\
\eq
and
 \bq \mu = \frac{\mathbfit{k}\cdot\mathbfit{r}}{k r} \ . \eq
The parallel part of a mode is given by $k_{\shortparallel} = \mu k$ and the perpendicular one by $k_{\perp} = k\sqrt{1-\mu^2}$.

The first term in \autoref{eq:P_g-def} arises from the Alcock-Paczinsky effect due to the relation between the Fourier modes of real space and those in the fiducial space. If we denote the values $\hat{H}$ and $\hat{D}$ as the quantities in the real or true cosmology, and the values $H$ and $D$ as corresponding to the fiducial cosmology, we obtain:
\bq
f_{\text{AP}}(z) = \frac{D_A^2 \hat{H}}{\hat{D}_A^2 H} \ .
\eq

The second term in \autoref{eq:P_g-def} is due to the limited resolution of any telescope, which means that the observed small-scale perturbations are suppressed. Assuming Gaussian errors $\sigma_{\shortparallel}(z)$ and $\sigma_{\perp}(z)$ on coordinates parallel and perpendicular to the line of sight at redshift $z$, respectively, the suppression factor is:
\bq
f_{\text{res}}(k,\mu,z) = \exp\left(-k^2\left[\mu^2\cdot\left(\sigma_{\shortparallel}^2(z)-\sigma_{\perp}^2(z)\right)+\sigma_{\perp}^2(z)\right]\right) \ .
\eq
The suppression factor is independent of the fiducial cosmology.

In any galaxy observation, there are additional sources of redshift alongside the cosmological redshift. The classical Doppler effect due to the velocity of galaxies produces an apparent anisotropy in the redshift-space power spectrum. The redshift-space distortion effects are parametrized by the third term in \autoref{eq:P_g-def}:
\bq
\label{eq:RSD}
f_{\text{RSD}}(\hat{k},\hat{\mu},z) = \left( 1+\beta(\hat{k},z) \, {\hat{\mu}}^2 \right)^2 e^{-{\hat{k}}^2{\hat{\mu}}^2\sigma_{\text{NL}}^2} \ .
\eq

The first term in brackets corresponds to the Kaiser formula \citep{Kaiser1987}. The term $\beta$ is the growth rate $f(\hat{k},z)$ corrected by the galaxy bias $b(z)$:
\begin{align}
\label{eq:beta}
\beta(\hat{k},z)& \equiv  \frac{f(\hat{k},z)}{b(z)  \nonumber}\\
& \equiv \frac{1}{b(z)} \frac{\dd \ln\left(\sqrt{P_\mathrm{m}(\hat{k},z)}\right)}{\dd \ln a} \nonumber \\
&= -\frac{1+z}{2b(z)} \frac{\dd \ln P_\mathrm{m}(\hat{k},z)}{\dd z} \ .
\end{align}
The bias relates the density perturbations in the galaxy field $\delta_\mathrm{g}$ to the dark matter density perturbations $\delta_\mathrm{m}$. We assume a linear approximation where the bias is scale-independent, so that:
\bq
\delta_\mathrm{g} = b(z)  \delta_\mathrm{m} \ .
\eq

The exponential term in \autoref{eq:RSD} accounts for the elongation in redshift-space maps along the line of sight within overdense regions, known as the `Fingers of God' effect. We include the term $\sigma_\text{NL}$ as a nuisance parameter in our forecasts, with a fiducial value of $7$~Mpc, and a prior range from $4$ to $10$~Mpc.

Our galaxy clustering survey is divided into redshift bins of width $\Delta z=0.1$ with mean redshift $\bar{z}$. 
The experimental shot noise in each redshift bin is parametrized by: 
\bq
P_{N}(\bar{z}) = \frac{1}{\bar{n}(\bar{z})} = \frac{V_r(\bar{z})}{N(\bar{z})} \ ,
\eq
where $N(\bar{z})$ is the number of galaxies in the bin, $V_r(\bar{z})$ the volume of the bin and $\bar{n}(\bar{z})$ the galaxy number density. 
The volume of each redshift bin is:
\begin{align}
V_r(\bar{z}) &=  4\pi f_{\text{sky}} \int_{\Delta r(\bar{z})}r^2 { \dd}r \nonumber \\
 &=\frac{4\pi}{3} f_{\text{sky}} \left[r^3\left(\bar{z}+\frac{\Delta z}{2}\right)-r^3\left(\bar{z}-\frac{\Delta z}{2}\right)\right] \ ,
\label{eq:Vr}
\end{align}
where $f_{\text{sky}}$ is the fraction of the sky covered by the survey.

We include massive neutrinos in our cosmological models. In such models, the clustering of halos is determined by cold dark matter and baryons only, and not by massive neutrinos. The calculation of the observed galaxy power spectrum in \citet{Sprenger2018} used here accounts for this by ignoring the contribution of light massive neutrinos with a free-streaming length larger than the typical size of a galaxy. It therefore includes only the cold dark matter and baryon field $P_\mathrm{cb}$, rather than the full matter field $P_\mathrm{m}$ (i.e. cold dark matter, baryons, and massive neutrinos) in \autoref{eq:P_g-def}, so that the equation is modified to:
\begin{align}
\label{eq:P_g_def_cb}
P_{\mathrm{g}}(k,\mu,z) = & f_{\text{AP}}(z) \times f_{\text{res}}(k,\mu,z) \nonumber \\
& \times f_{\text{RSD}}(\hat{k},\hat{\mu},z) \times b^2(z) \times P_{\mathrm{cb}}(\hat{k},z) \ .
\end{align}

The $\beta$ term of the Kaiser formula in \autoref{eq:beta} is then modified to:
\bq
\beta(\hat{k},z) = -\frac{1+z}{2b(z)} \frac{\dd \ln P_\mathrm{cb}(\hat{k},z)}{\dd z} \ ,
\eq
with the bias now defined as  $\delta_\mathrm{g} = b(z) \times \delta_\mathrm{cb}$.

Putting all this together, we finally obtain the observed galaxy clustering power spectrum in each bin:
\bq
\label{eq:pobs}
P_\mathrm{obs}(k,\mu,\bar{z}) = P_\mathrm{g}(k,\mu,\bar{z}) + P_{N}(\bar{z}) \ .
\eq

\subsubsection{Galaxy clustering likelihood}
\label{sec:GC_likelihood}

Since we are dealing with simulated data, $P_\mathrm{obs}$ is either produced by our fiducial cosmology, or by the points sampled in parameter space. If we denote each quantity by the label $f$ (fiducial cosmology), or $s$ (sample cosmology), the galaxy clustering likelihood can be written as:
\begin{eqnarray}
\label{eq:GC_likelihood}
\chi^2 = & \sum_{\bar{z}}\int (k^f)^2\dd k^f \int_{-1}^{1} \dd \mu^f \frac{V_r^f}{2(2\pi)^2}   \nonumber \\ 
& \times \left[\frac{\frac{H^f }{(D_A^f)^2}P^f_\mathrm{g}(k^f,\mu^f)-\frac{H^s }{(D_A^s)^2} P^s_\mathrm{g}(k^s,\mu^s)}{\frac{H^s }{(D_A^s)^2}P^s_\mathrm{g}(k^s,\mu^s)+\frac{H^f }{(D_A^f)^2}\frac{V_r^f}{N}}\right]^2 \ ,
\end{eqnarray}
where $\chi^2=-2 \ln \mathcal{L}$.

\subsubsection{\textit{Euclid} galaxy clustering specifications}

We use a redshift range from $0.7$ to $2.0$, which is the approximate range accessible to \textit{Euclid}. For the bin centres $\bar{z}$, we use a minimum redshift $z_\mr{min}=0.75$ and a maximum redshift $z_\mr{max}=1.95$, with the entire redshift range divided into $13$ bins. 

The error on spectroscopic redshift measurements is assumed to be $\sigma_z = 0.001(1+z)$. The effect of angular resolution is neglected, so that $\sigma_{\perp}$ is set to 0. The radial distance error is a function of the redshift error, and is cosmology-dependent:
\bq
\sigma_{\shortparallel} = \frac{c}{H} \sigma_z \ .
\eq
The galaxy number count distribution $\dd N(z) / \dd z$ per $\text{deg}^2$ assumes a limiting flux of $3 \times 10^{-16}$\,erg\,s$^{-1}$\,cm$^{-2}$, and is taken from \citet{Pozzetti2016}.

We use a sky fraction of $f_{\text{sky}}=0.3636$. The total number of detected galaxies in a given redshift bin is:
\bq
N(\bar{z}) = 41253 f_{\text{sky}} \text{deg}^2 \int_{\bar{z}-\frac{\Delta z}{2}}^{\bar{z}+\frac{\Delta z}{2}}\frac{\dd N(z)/\dd z}{1\text{deg}^2}\dd z \ .
\eq
The bias factor corresponding to galaxies detected by \textit{Euclid}  is assumed to be close to the simple relation:
\bq
b(z) = \sqrt{1+z} \ ,
\eq
which is also used in \citet{Audren2013}.

Two parameters introduced in \citet{Sprenger2018} account for inaccuracies in this relation:
\bq
b(z) = \beta_0^{\text{Euclid}}(1+z)^{0.5\beta_1^{\text{Euclid}}} \ .
\eq
We include $\beta_0^{\text{Euclid}}$ and $\beta_1^{\text{Euclid}}$ in our MCMC simulations as nuisance parameters. We assign a mean value of $1$ to both, with an unbounded prior range.

\subsection{The non-linear theoretical uncertainty}

With its wide sky coverage and its catalogue of billions of stars and galaxies, \textit{Euclid} will be a great leap forward in observational cosmology. However, this will significantly increase the size of the sampling variance and shot noise compared to current surveys. The leading source of error on small scales, which then impacts parameter extraction and model selection, will be theoretical errors.

In \citet{Sprenger2018}, a novel method was introduced for dealing with the theoretical uncertainties, which we use here. This strategy defines a cutoff $k_\mathrm{NL}$. All theoretical uncertainties up to this wavenumber are ignored, while all the information above it is discarded. The redshift dependence of non-linear effects is parametrized as:
\bq
\label{eq:nonlinear}
k_\mathrm{NL}(z)=k_\mathrm{NL}(0) (1+z)^{2/(2+n_\mathrm{s})} \ .
\eq

Our use of this parametrization with cosmological models where the tilt $n_\mathrm{s}$ is not explicitly present is justified by two considerations. First, the envelope of the wiggles on the Wiggly Whipped primordial power spectrum corresponds to the tilt on the power-law spectrum. Our parameter choices ensure that the resulting spectrum is compatible with the current best data from \textit{Planck}. Secondly, the wiggles die down well before the non-linear part of the spectrum (see \autoref{fig:pk}). 

In \citet{Sprenger2018} two frameworks for modelling the theoretical error are defined. The first is a `realistic' case where the parametrization of the error is trusted up to large wavenumbers. The information from small scales is gradually suppressed by a growing relative error function. The second is a `conservative' case where the same error function is used with a sharp cut-off. 

The parameters for our galaxy clustering forecast correspond to the `conservative' setup. We adopt a cut-off on large wavelengths at $k_\mr{min}=0.02~\Mpc$. This eliminates scales which are bigger than the bin width or which violate the small-angle approximation. On small wavelengths, we use a theoretical uncertainty with $k_\mr{NL}(0)=0.2 \hpM$.

Similarly, we adopt the `conservative' setup for the cosmic shear forecast. We include multipoles from $\ell_\mathrm{min}=5$ up to a bin-dependent non-linear cut-off determined by $k_\mr{NL}(0)=0.5 \hpM$.

\subsection{Cosmology}
\label{sec:Cosmology}

In this paper, we work within the framework of Friedmann-Robertson-Walker cosmology, and we assume a flat spatial geometry for all our models. We consider two main classes of cosmological models. The first is $\LCDM$ with a power-law primordial power spectrum, or the Concordance Model of Cosmology. The second class consists of Wiggly Whipped Inflation models. These two classes are distinguished by the shape of the primordial power spectrum. For the former, they are featureless. For the latter, they contain features.

The background $\LCDM$ cosmology for all the models contains baryonic and cold dark matter, massive neutrinos, and a cosmological constant or constant dark energy. This is parametrized by: the baryon density $\omega_{\mathrm{b}}=\Omega_\mathrm{b} h^2$, the cold dark matter density $\omega_{\mathrm{cdm}}=\Omega_\mathrm{cdm} h^2$, the Hubble parameter via the peak scale parameter $100\theta_{\mathrm{s}}$, and the optical depth to reionization $\tau_{\mathrm{reio}}$. 
We use the following fiducial values for the background cosmology to generate the mock data for all our models:
 $\omega_{\mathrm{b}}= 2.21\times 10^{-2}$,
 $\omega_{\mathrm{cdm}}=0.12$,
$100\theta_{\mathrm{s}}=1.0411$, and 
$\tau_{\mathrm{reio}}=0.09$.
We assume three neutrino species, with the total neutrino mass split according to a normal hierarchy. We therefore have 2 massless and 1 massive neutrinos. We keep all the values of the neutrino parameters fixed as follows: the sum of the neutrino masses $M_\mr{total}=0.06$~eV; and the number of effective neutrino species in the early Universe $N_\mr{eff}=3.046$. Our choice of neutrino fiducial values is motivated by the latest data from neutrino oscillation experiments \citep{NeutrinoHierarchy}, which show strong statistical support for a normal hierarchy.
Our Concordance Model is parametrized by two additional parameters for the power-law  primordial power spectrum: the scalar amplitude $A_{\mathrm{s}}$ and the scalar spectral index $n_{\text{s}}$. We use the following fiducial values: $\ln(10^{10}A_{\text{s}}) =3.0447$, and $n_{\text{s}}=0.9659$, with the pivot scale $k_0$ fixed at $0.05\, \Mpc$.

\begin{figure*}
	\includegraphics[width=\columnwidth]{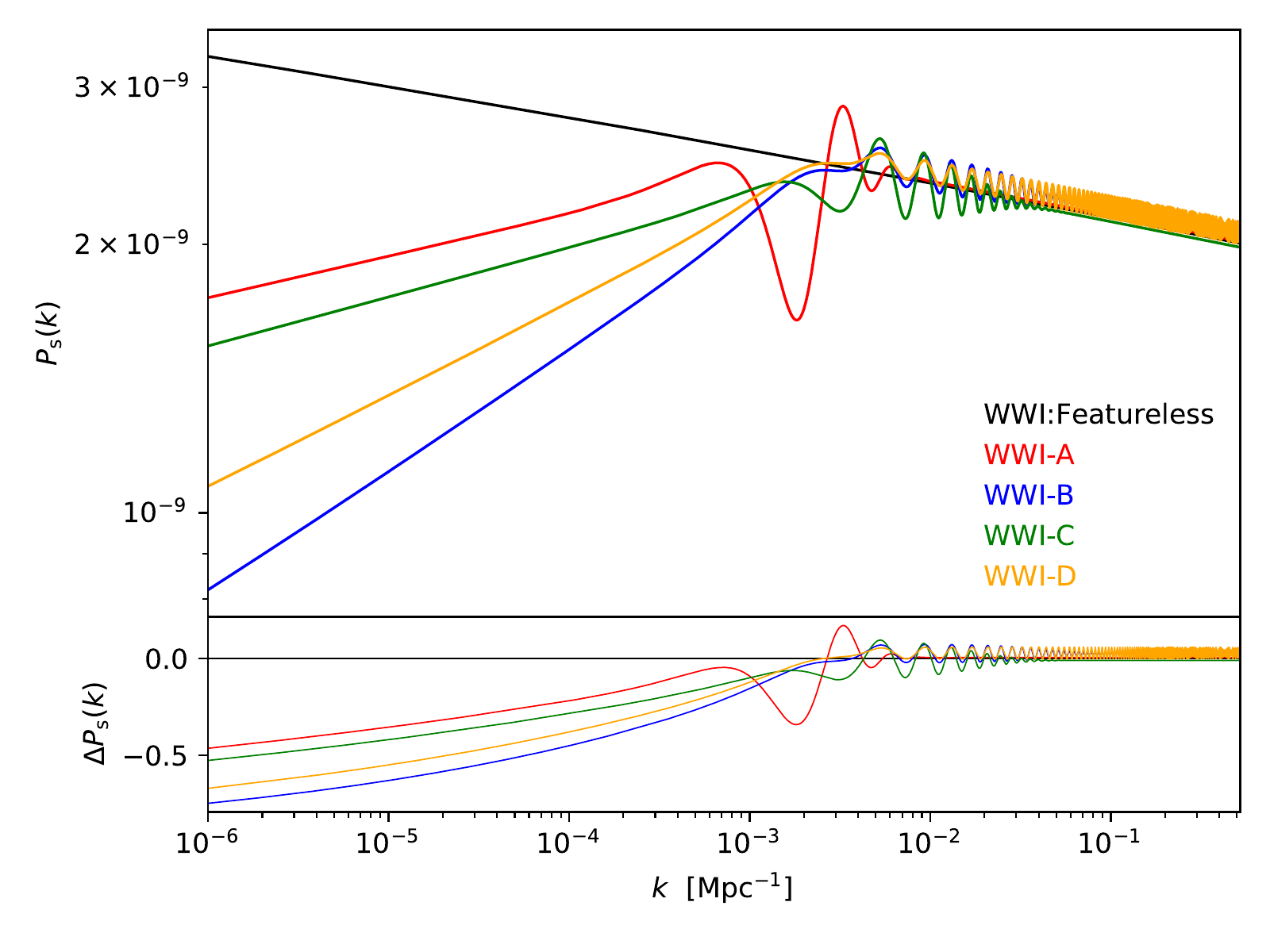}
	\includegraphics[width=\columnwidth]{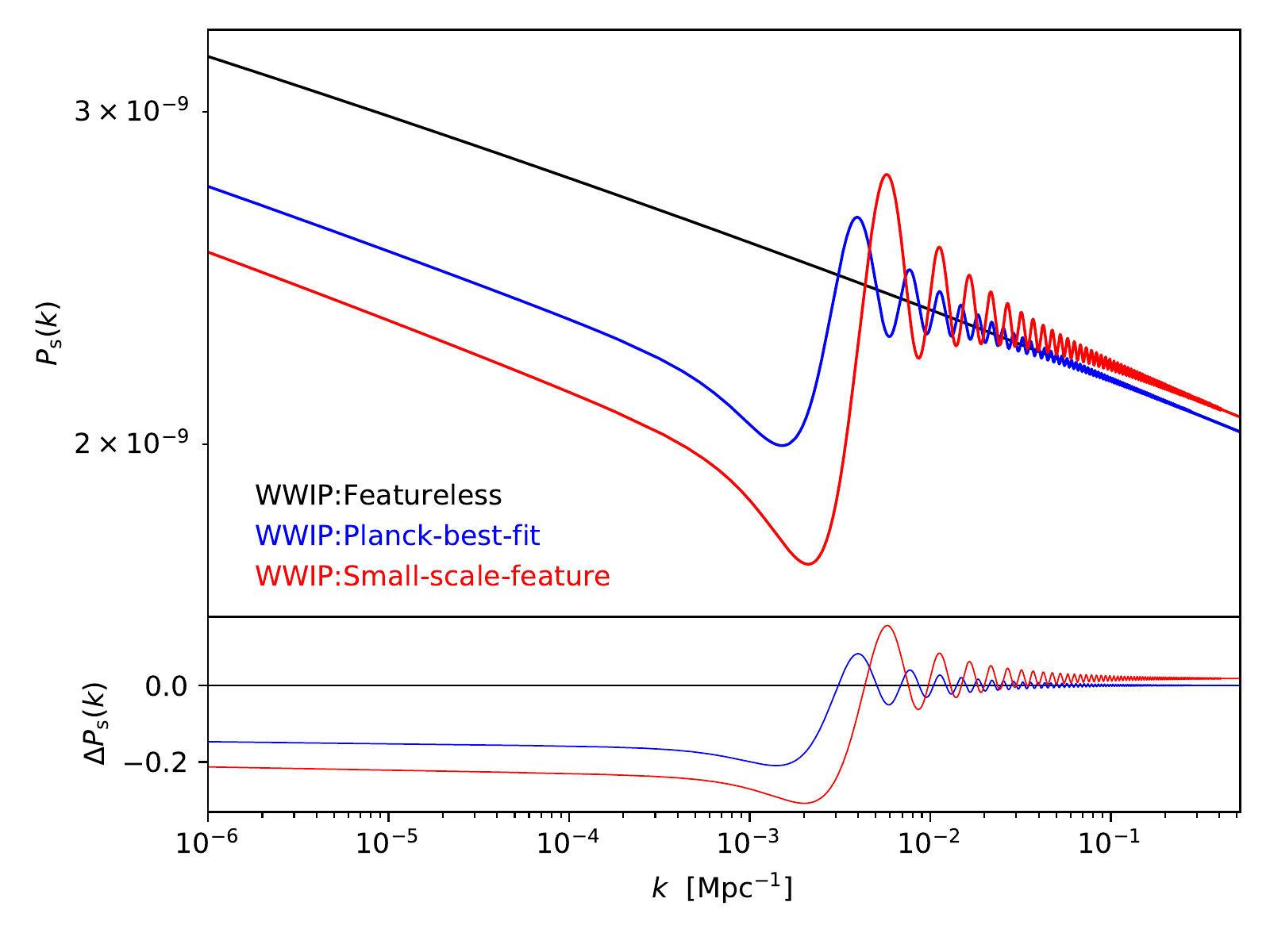}
    \caption{The scalar primordial power spectrum for the fiducial models considered in this paper. 
    The left and right panels show WWI and WWIP, respectively. The inset bottom panel in each plot shows the amplitude of the features relative to the featureless spectrum $P_0$ (i.e. $\Delta P(k)=P_0(k)-P(k ))/P_0(k)$).
 }
    \label{fig:pk_prim}
\end{figure*}

\begin{figure*}
	\includegraphics[width=\columnwidth]{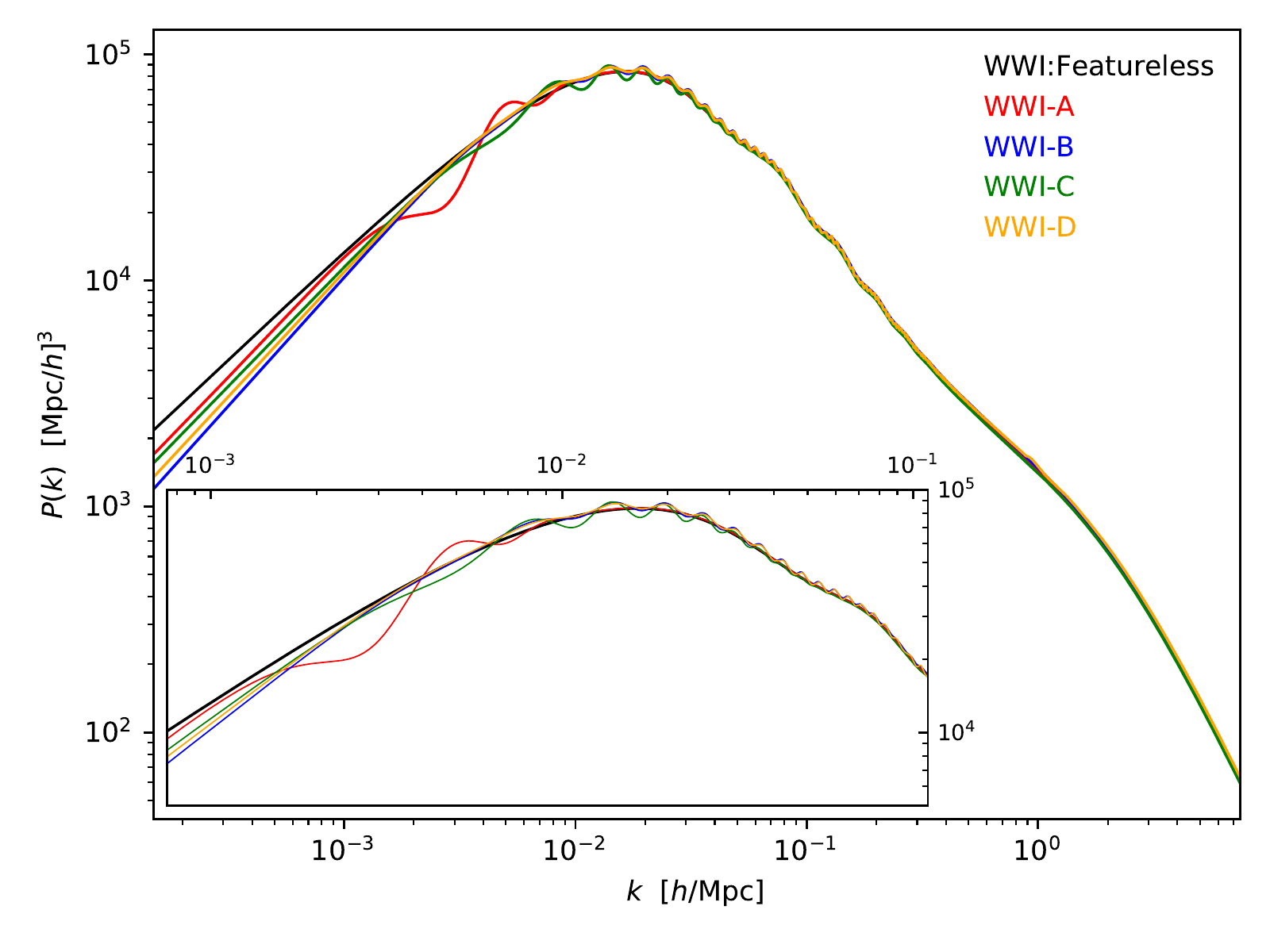}
	\includegraphics[width=\columnwidth]{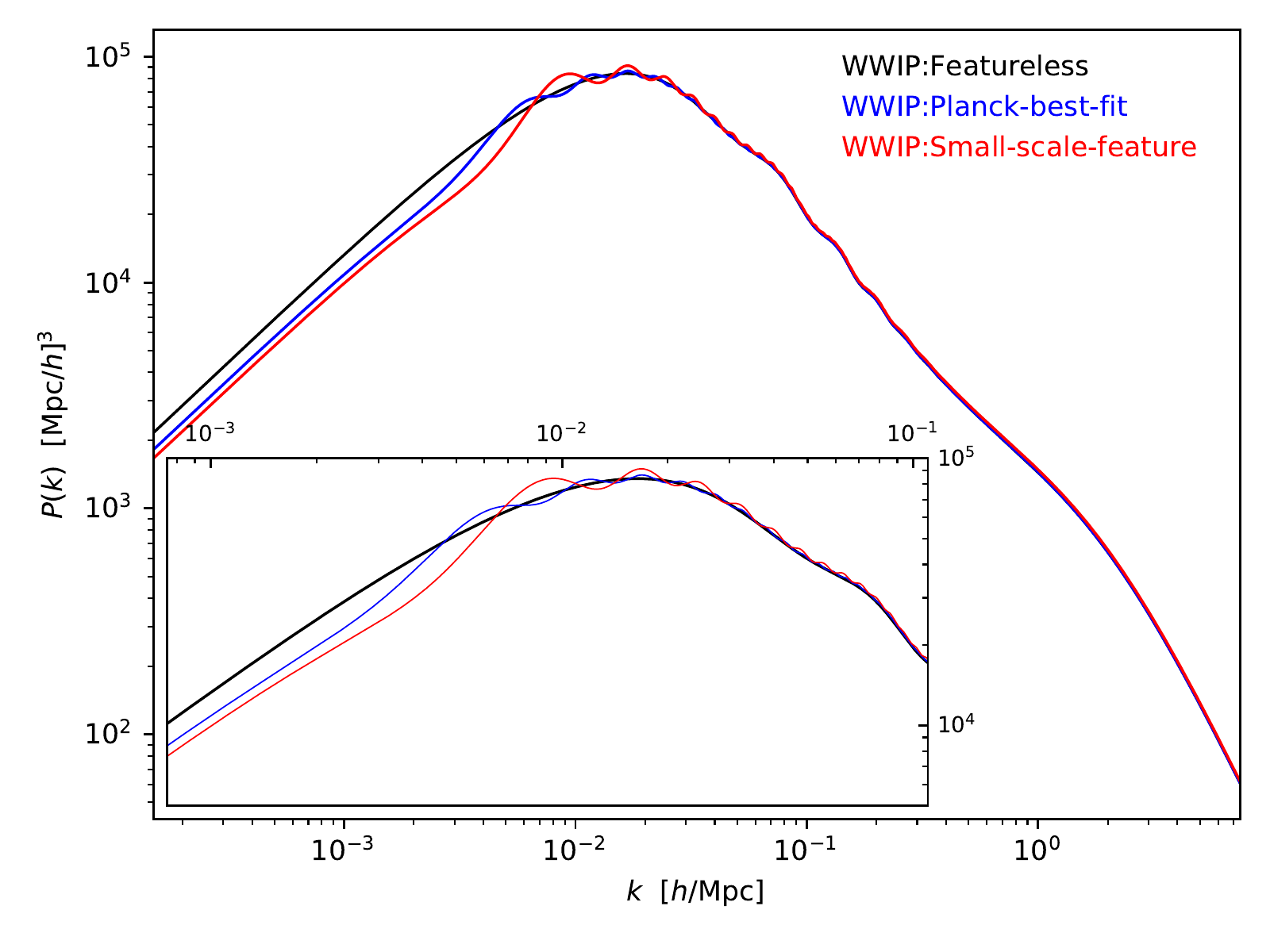}
    \caption{The full non-linear matter power spectrum at $z=0$ for the fiducial models considered in this paper. 
    The left and right panels show WWI and WWIP, respectively. The inset plots show the range from $k=0.001$ to $0.1$.}
    \label{fig:pk}
\end{figure*}

\begin{figure*}
	\includegraphics[width=\columnwidth]{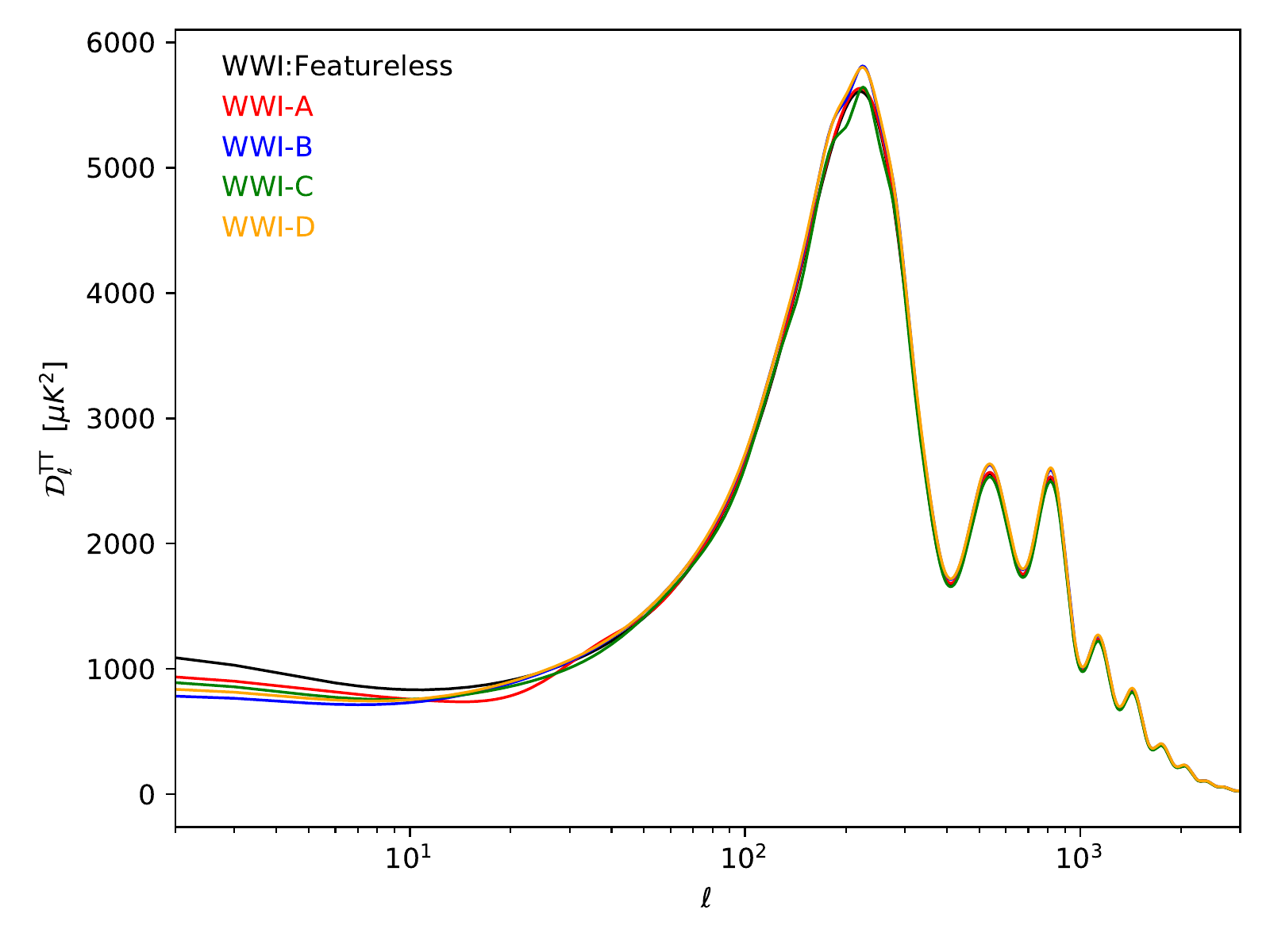}
	\includegraphics[width=\columnwidth]{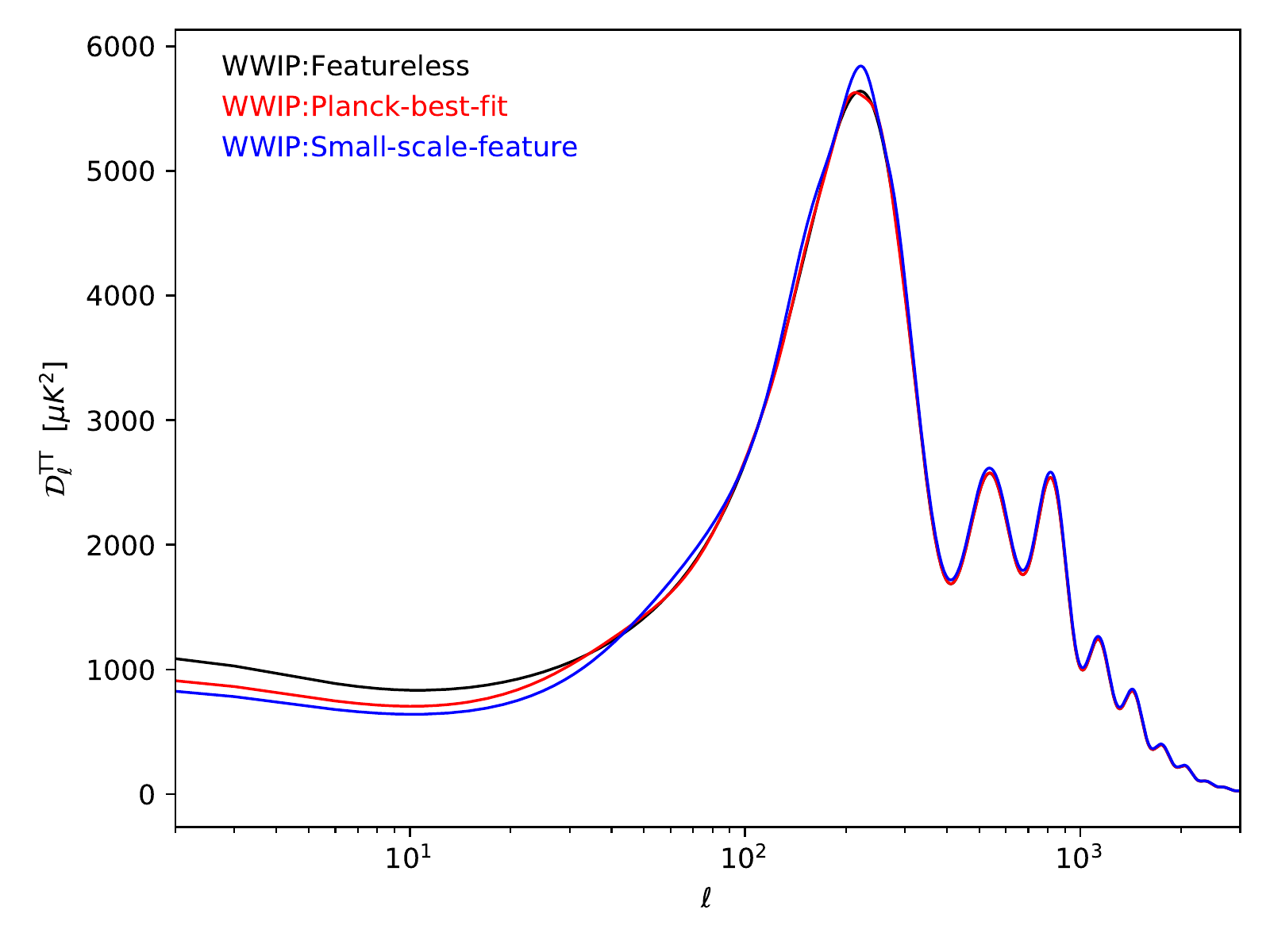}

	\includegraphics[width=\columnwidth]{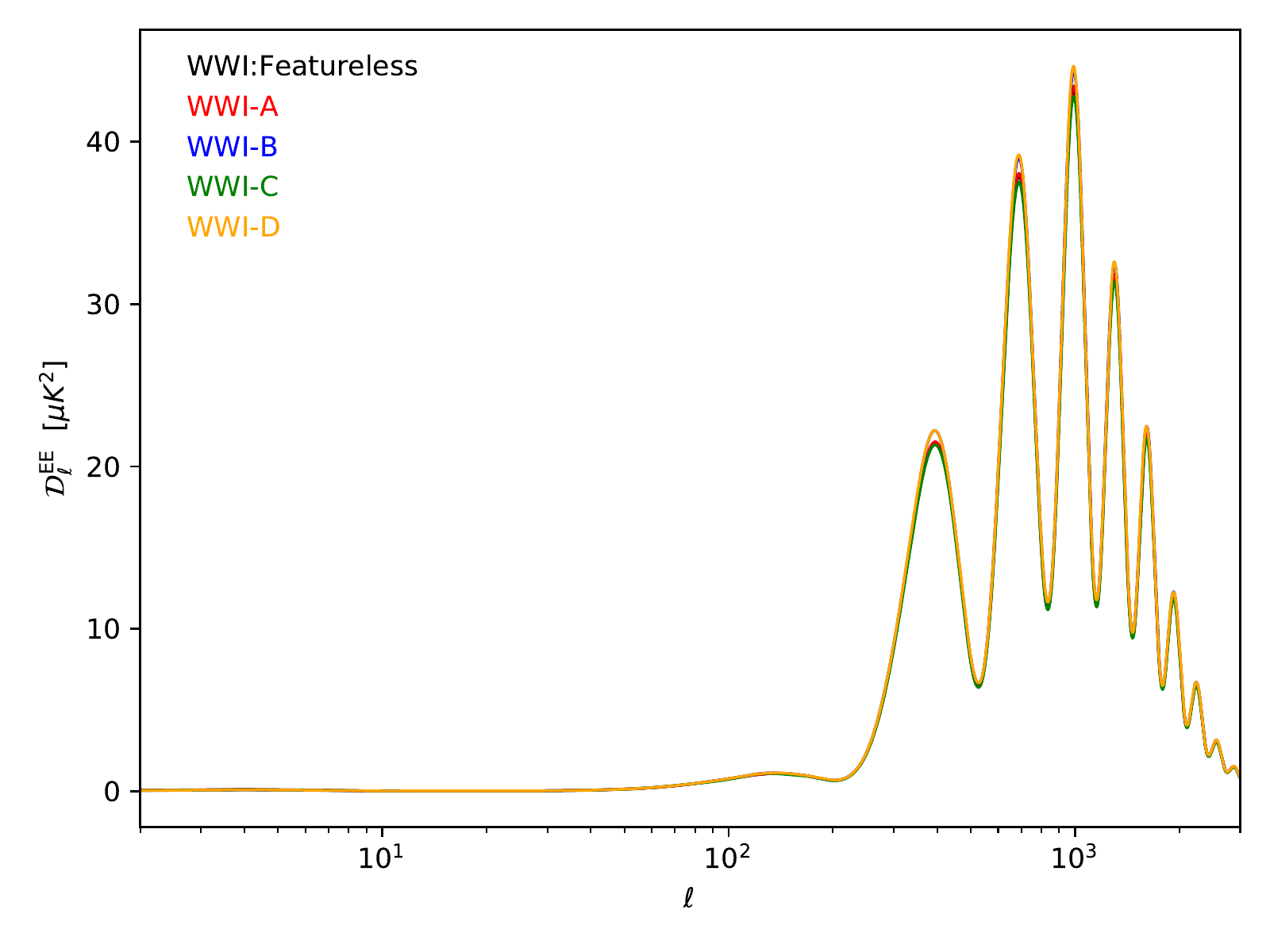}
	\includegraphics[width=\columnwidth]{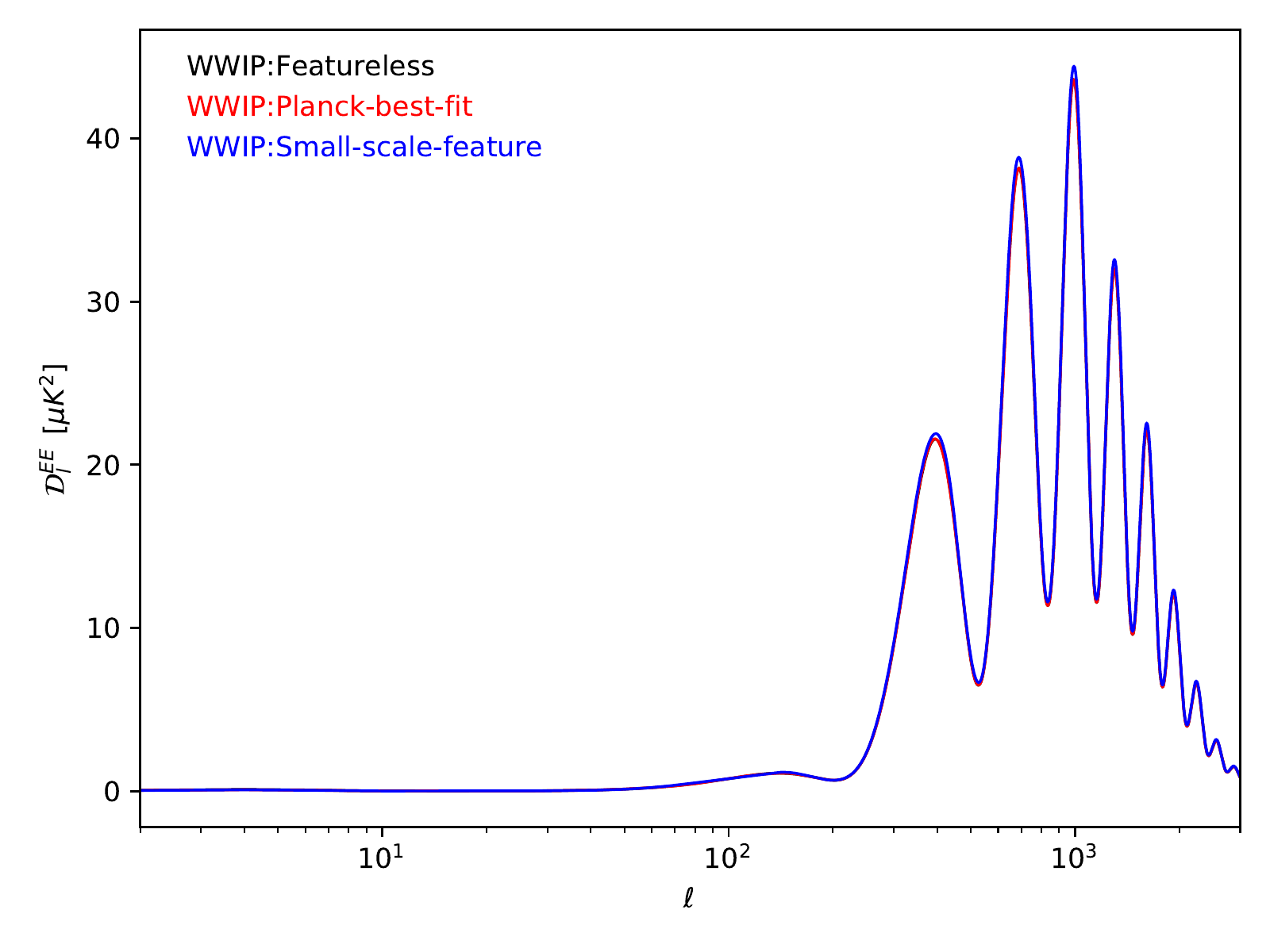}	
    \caption{The $TT$ (top) and $EE$ (bottom) CMB angular power spectrum for the fiducial models considered in this paper. 
    The left and right panels show WWI and WWIP, respectively.}
    \label{fig:TTEE}
\end{figure*}

The second class of models contain features in the primordial power spectrum with $\LCDM$ as background cosmology. Here, instead of using power law spectrum we use the numerical solution to the Klein-Gordon and Mukhanov-Sasaki equations for background scalar field evolution and cosmological perturbations respectively. Therefore in addition to four parameter describing the $\LCDM$ background we have the inflationary potential parameters. The fiducial values for inflation potential parameters used to produce the mock data are shown in \autoref{tab:InflationModels}. For our MCMC simulations, the sampled data is parametrized by five free inflation parameters for WWI ($\ln(10^{10}V_0)$, $\phi_{0}$, $\gamma$, $\phi_\mathrm{T}$, $\ln\delta$), and three free inflation parameters for WWIP ($\ln(10^{10}V_0)$, $\phi_{0}$, $\phi_\mathrm{T}$. We consider five fiducial primordial spectra for the WWI potential and three for the WWIP potential. For the WWI potential, four of the fiducial power spectra contain different types of features at different cosmological scales which represent local and global best fits to the \textit{Planck} data. We call these WWI-[A, B, C, D], following the nomenclature in \citet{Hazra2016}. As explained in \citet{Hazra2016}, WWI-A and WWI-C are local best-fitting values to individual and joint BICEP2/Keck and \textit{Planck} 2015 data sets, while WWI-B and WWI-D are close to a global best fit for all data sets, in the sense that they provide an improved fit compared to the power-law model. For the WWIP potential, we use the \textit{Planck} global best-fitting~\citet{Hazra2016} spectrum (hereafter, WWIP:Planck-best-fit) and another spectrum found within the 95\% confidence limits of \textit{Planck} data (hereafter, WWIP:Small-scale-feature). This has wiggles extending to smaller scales where the overlap with upcoming \textit{Euclid} data is better. We also consider two spectra without features for both  WWI and WWIP. These two fiducial spectra are obtained by fixing $\phi_{0}=0$, $\gamma=0$ for WWI, and $\phi_{0}=0$ for WWIP, respectively. These two spectra are used as a null test. The primordial power spectra produced by the feature models are shown in \autoref{fig:pk_prim}. 
Corresponding matter power spectra for these fiducial potential parameters are provided in \autoref{fig:pk} and the CMB temperature and polarization angular power spectra are provided in \autoref{fig:TTEE}. Since these fiducials represent features at different scales and amplitude, large-scale-structure data will have different constraining power when combined with CMB from \textit{Planck}. Using the sensitivity of the cosmic-svariance-limited proposed CMB polarization survey CORE~\citep{CORE:params,CORE:inflation}, forecasts on these types of features were performed in~\citet{Hazra:coreforecast}~\footnote{Using N-body simulations, the effects of some of these features have been tested in~\citet{LHuillier:2017lgm} and a few other features have been tested in~\citet{Ballardini:2020}.}. It was shown that while the largest-scale features ($\ell<50$) cannot be detected with next-generation CMB surveys beyond 95\% C.L., intermediate and small-scale oscillations can be discovered with high statistical significance. Since the proposed CORE mission was not approved we expect that a joint combination of \textit{Euclid} and \textit{Planck} can identify certain types of features if they represent the {\it true} model of the Universe and fall within the good signal-to-noise region of both the surveys.

\begin{table*} \begin{tabular}{l c c c c c}
\hline
Model & $\ln(10^{10}V_0)$ & $\phi_{0}$ & $\gamma$ & $\phi_\mathrm{T}$ & $\ln\delta$ \\
\hline
WWI:Featureless & 1.73  & 0 & 0 & -- & -- \\ 
WWI--A           & 1.73  & 0.0137 & 0.019 & 7.89 & -4.5 \\
WWI--B           & 1.75  & 0.0038 & 0.04 & 7.91 & -7.1 \\
WWI--C           & 1.72  & 0.0058 & 0.02 & 7.91 & -6 \\
WWI--D           & 1.76  & 0.003 & 0.033 & 7.91 & -11 \\
WWIP:Featureless & 0.282     & 0   & -- & -- & -- \\
WWIP:Planck-best-fit & 0.282     & 0.11 & -- & 4.51 & -- \\
WWIP:Small-scale-feature & 0.3 & 0.18 &-- & 4.5 & -- \\
\hline
\end{tabular}
\caption{
Fiducial values: Inflationary potential parameters used to obtain the fiducial primordial power spectrum. We have used two types of potential in this framework: WWI (see \autoref{eq:equation-WWI}) and WWIP (see \autoref{eq:equation-WWIP}). For both inflation models, we include a featureless case (labelled WWI:Featureless and WWIP:Featureless, respectively). WWI--A, B, C and D and WWIP:Planck-best-fit represents the best fit potential parameters to the combined \textit{Planck} temperature and polarization data. WWIP:Small-scale-feature corresponds to a particular power spectrum that has features at small scales (within \textit{Planck} 95\% confidence limits), which ensures better overlap with cosmological scales to be probed by \textit{Euclid}.
}
\label{tab:InflationModels}
\end{table*}

In order to compute the primordial power spectrum from the inflation models, we use the \textsc{bingo} package \citep{Bingo}\footnote{BI-spectra and non-Gaussianity operator. Available on \href{https://github.com/dkhaz/bingo}{https://github.com/dkhaz/bingo}}. Since the existence of fast-roll limits the use of analytical approximations in obtaining the power spectrum, \textsc{bingo} is a necessary tool. We solve both the cosmological background and perturbation equations during inflation using adaptive stepsize, and adapt \textsc{bingo} to output the primordial power spectrum directly as an input for \textsc{class} via an external command.
We evaluate the sensitivity to cosmological parameters of \textit{Euclid} combined with \textit{Planck} by performing MCMC forecasts with the joint data, and compare these against \textit{Planck}-only constraints. We generate spectra for the fiducial models, which we then use as our mock data. This method has two advantages over the Fisher matrix formalism. First, it avoids the problem of numerical instabilities, particularly those linked to the choice of step size for the numerical derivatives. Secondly, it allows us to work with non-Gaussian errors, especially those which may arise with non-standard cosmologies such as ours. 
We use the MCMC sampler \textsc{montepython} \citep{MontePython3} with the Boltzmann solver \textsc{class} \citep{CLASS} to generate MCMC samples using a Metropolis-Hasting algorithm. Since we do not yet have any data from \textit{Euclid}, we use data generated from a fiducial cosmological model. We include both cosmic shear and galaxy clustering, using the likelihoods from \citet{Sprenger2018}.
We include the non-linear part of the power spectrum in our forecasts. The non-linear contribution is calculated within \textsc{class} using the \textsc{halofit} \cite{Takahashi2012,Bird2012} semi-analytical formula.

\section{Results}
\label{sec:Results}

In this section, we present our results.
We fit the theoretical sampled angular power spectra and matter power spectra to the fiducial mock data of the corresponding models. 
Using the MCMC technique, we obtained \textit{Planck}-only (labelled '\textit{Planck}' and joint \textit{Euclid}+\textit{Planck} constraints, which are presented in this section. \textit{Euclid} here includes galaxy clustering and cosmic shear. For convenience, we just use the label `\textit{Euclid}'. For both \textit{Planck}-only and \textit{Euclid}+\textit{Planck}, we keep the same mock data in order to have a consistent comparison. The MCMC chains were analysed using \textsc{getdist}.

\begin{figure*}
	\includegraphics[width=168mm]{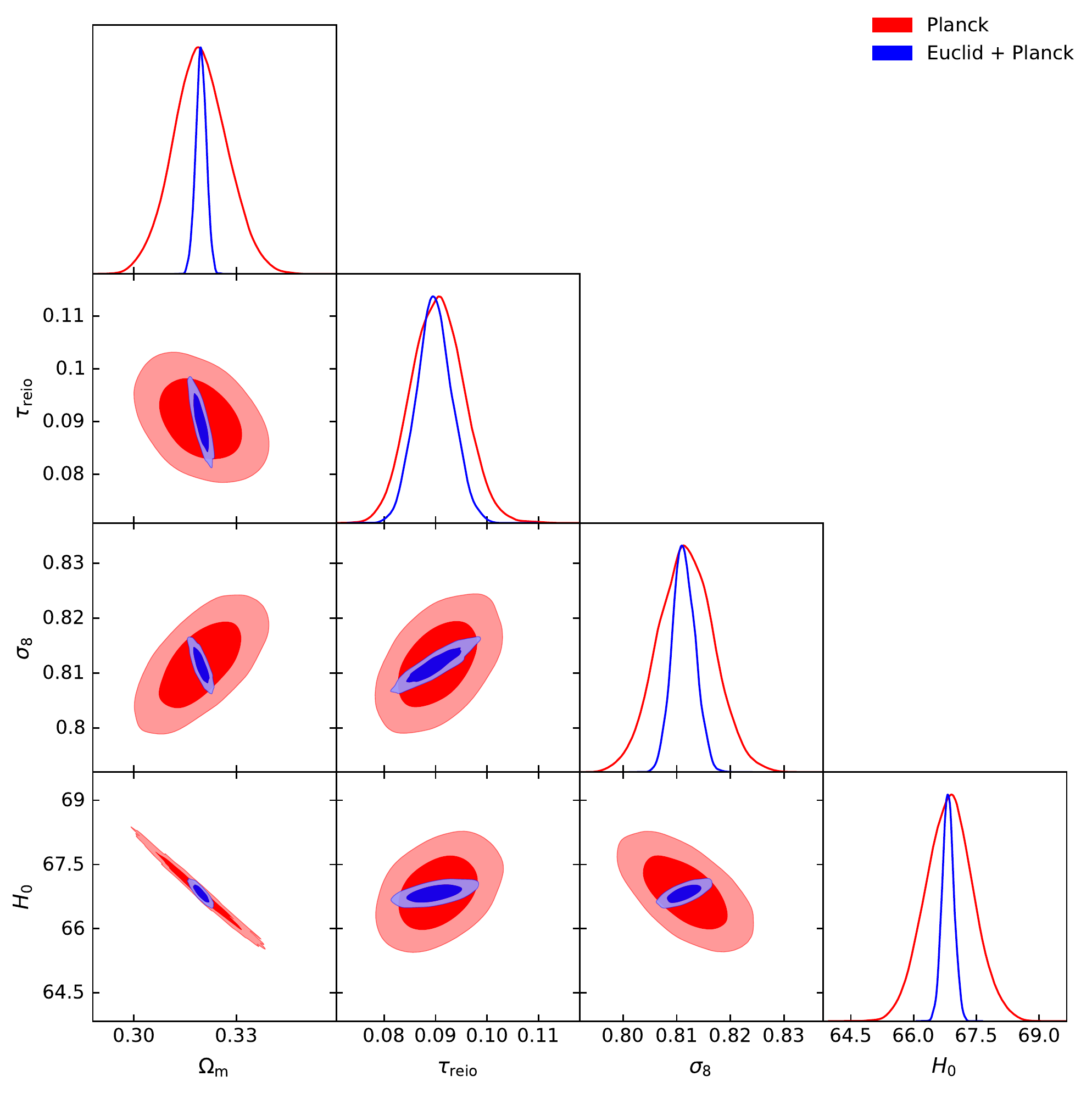}
    \caption{One-dimensional posteriors and marginalized $68$ per cent and $95$ per cent contours for \textit{Planck} (red) and \textit{Euclid} galaxy clustering $+$ \textit{Euclid} cosmic shear $+$  \textit{Planck} (blue) for four parameters in the $\LCDM$ Concordance Model. Constraints on the Hubble parameter are
    improved by \textit{Euclid} due to its degeneracy with $\Omega_\mathrm{m}$, which is evident in this plot.}
    \label{fig:LCDM_baseparameters}
\end{figure*}

In \autoref{tab:LCDM_table} we provide the 
projected constraints on the $\LCDM$ model with power law primordial spectrum (i.e. the Concordance Model) with both dataset combinations. Note that first six parameters in the table are the parameters used for MCMC analysis. We include four derived parameters: the dark energy density $\Omega_{\Lambda }$, the matter density $\Omega_\mathrm{m }$, the Hubble constant $H_0$, and the power spectrum normalization parameter $\sigma_8$, defined as the root-mean-square amplitude of the density contrast inside an $8\, h^{-1}\text{Mpc}$ sphere. 
We also plot the one-dimensional posteriors on  
four parameters ($\Omega_{\mathrm{m}}$, $\tau_{\mathrm{reio}}$, $\sigma_{8}$ and $H_0$) and their marginalized contours in \autoref{fig:LCDM_baseparameters}. The table reflects improvements in all the parameter constraints when \textit{Euclid} mock likelihood is combined with CMB. While the baryon density experiences a marginal improvement, constraints on CDM density becomes 4 times tighter which is reflected in the posterior of matter density. Constraints on $H_0$ are improved through its degeneracies with other parameters, especially $\Omega_\mathrm{m}$. The amplitude ($\ln[10^{10} A_{\mathrm s}]$) and tilt ($n_\mathrm{s}$) of the primordial power spectrum are expected to be constrained 30-40 per cent better with \textit{Euclid} compared to the present bounds. Since \textit{Euclid} will probe small scales comparatively better than \textit{Planck} and the weak lensing will probe order-of-magnitude smaller scales beyond the \textit{Planck} CMB probed scales, the long lever arm on the two-point correlations at small scales will be able to improve the constraints. The $\sigma_8$ being the integral of the matter power spectrum (which, in turn, is defined by the primordial spectrum amplitude and tilt and the transfer function) it is also expected to be constrained two-fold tighter than present constraints from \textit{Planck}. The optical depth is not directly associated with the physical processes probed by \textit{Euclid}. However, since the amplitude of the primordial spectrum is completely degenerate with optical depth, improvement in the constraints on the amplitude also improves the constraints on the $\tau_{\mathrm reio}$. Note that here the mean value of $\tau_{\mathrm reio}$ is higher than the recently released \textit{Planck}value that is obtained simply as an artefact of using higher $\tau_{\mathrm reio}$ in the fiducial cosmology. Use of higher $\tau_{\mathrm reio}$ does not affect our analysis as \textit{Euclid} cannot directly constrain optical depth and in the forecast we are concerned only about the bounds on the parameters, and not their mean value.

\begin{table}
	\centering
	\caption{Forecast $68\%$ confidence intervals ($1\sigma$) for the $\LCDM$ Concordance Model using \textit{Planck} alone, and joint \textit{Euclid}  + \textit{Planck} data. The first six parameters define the cosmological model. The last four are derived parameters.}
	\label{tab:LCDM_table}
	
\begin{tabular} { l c c}
\hline
 Parameter &  Planck & Euclid+Planck\\
\hline
{\boldmath$10^{-2}\omega_\mathrm{b }$} & $2.210\pm 0.015            $ & $2.210^{+0.011}_{-0.012}   $\\
{\boldmath$\omega_\mathrm{cdm }$} & $0.1200\pm 0.0013          $ & $0.12000\pm 0.00031        $\\
{\boldmath$100\theta_\mathrm{s }$} & $1.04110\pm 0.00035        $ & $1.04108\pm 0.00031        $\\
{\boldmath$\tau_\mathrm{reio }$} & $0.0905\pm 0.0051          $ & $0.0898\pm 0.0035          $\\
{\boldmath$\ln10^{10}A_\mathrm{s }$} & $3.0455\pm 0.0096          $ & $3.0444\pm 0.0061          $\\
{\boldmath$n_{\mathrm{s} }         $} & $0.9660\pm 0.0037          $ & $0.9658\pm 0.0022          $\\
{\boldmath$\Omega_{\Lambda }$} & $0.6805\pm 0.0080          $ & $0.6803\pm 0.0016          $\\
{\boldmath$\Omega_\mathrm{m }$} & $0.3194\pm 0.0080          $ & $0.3196\pm 0.0016          $\\
{\boldmath$H_0            $} & $66.85\pm 0.57             $ & $66.82^{+0.13}_{-0.15}     $\\
{\boldmath$\sigma_8       $} & $0.8117\pm 0.0052          $ & $0.8114\pm 0.0021          $\\
\hline
\end{tabular}

\end{table}

\begin{table}
	\centering
	\caption{$1 \sigma$ confidence intervals for cosmological parameters with WWI:Featureless as the fiducial cosmology. 
	}
	\label{tab:WWIFeatureless}
	
\begin{tabular} { l c  c}
\hline
 Parameter &  Planck & Euclid+Planck\\
\hline

{\boldmath$10^{-2}\omega_\mathrm{b }$} 	&	 $2.210^{+0.015}_{-0.013}   $	&	 $2.209\pm 0.011            $	\\
{\boldmath$\omega_\mathrm{cdm }$} 	&	 $0.11997\pm 0.00091        $	&	 $0.11999\pm 0.00025        $	\\
{\boldmath$100\theta_\mathrm{s }$} 	&	 $1.04111\pm 0.00033        $	&	 $1.04111\pm 0.00031        $	\\
{\boldmath$\tau_\mathrm{reio }$} 	&	 $0.0906^{+0.0047}_{-0.0052}$	&	 $0.0902\pm 0.0029          $	\\
{\boldmath$\ln(10^{10}V_0)$} 	&	 $1.7313\pm 0.0098          $	&	 $1.7304\pm 0.0055          $	\\
{\boldmath$\phi_{0}       $} 	&	 $< 0.0195                  $	&	 $< 0.0207                  $	\\
{\boldmath$\gamma         $} 	&	 $< 0.0953                  $	&	 Unbounded                        	\\
{\boldmath$\phi_\mathrm{T}$} 	&	 $< 7.78                    $	&	 $< 7.78                    $	\\
{\boldmath$\ln\delta      $} 	&	 $-4.5^{+1.7}_{-1.1}        $	&	 $-4.54^{+1.7}_{-0.88}      $	\\
{\boldmath$\Omega_{\Lambda }$} 	&	 $0.6805\pm 0.0058          $	&	 $0.6804\pm 0.0013          $	\\
{\boldmath$\Omega_\mathrm{m }$} 	&	 $0.3194\pm 0.0058          $	&	 $0.3195\pm 0.0013          $	\\
{\boldmath$H_0            $} 	&	 $66.85\pm 0.42             $	&	 $66.83\pm 0.13             $	\\
{\boldmath$\sigma_8       $} 	&	 $0.8299\pm 0.0049          $	&	 $0.8297\pm 0.0018          $	\\

\hline
\end{tabular}

\end{table}

\begin{figure*}
	\includegraphics[width=168mm]{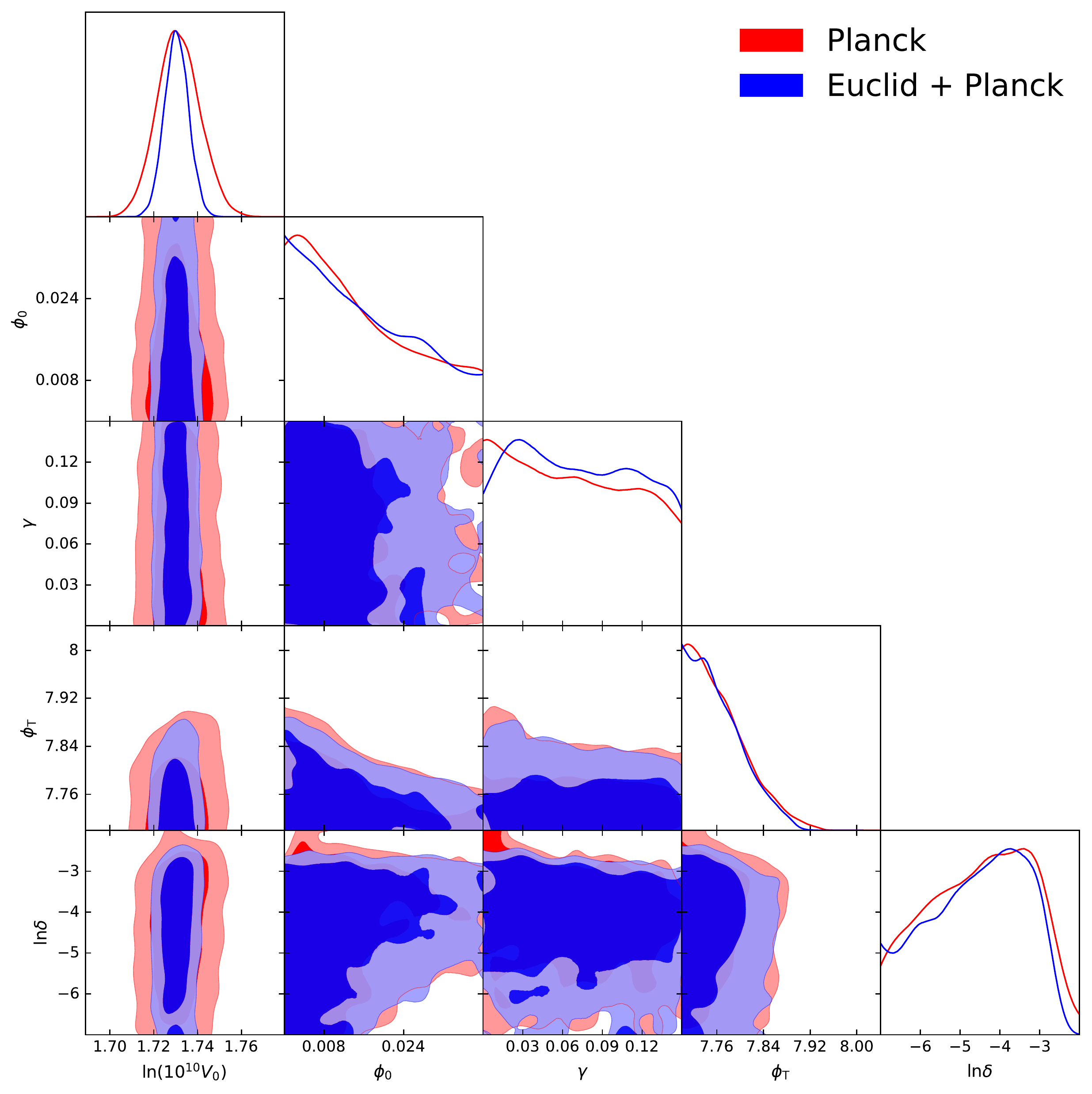}
    \caption{One-dimensional posteriors and marginalized contours for the inflation parameters in the WWI:Featureless model. 
    The addition of \textit{Euclid} data results in a significant improvement in constraints for the amplitude parameter.
   }
    \label{fig:WWIFeatureless}
\end{figure*}

\begin{table}
	\centering
	\caption{ $1 \sigma$ confidence intervals for cosmological parameters with WWI--A as the fiducial cosmology. }
	\label{tab:WWIA}
	
\begin{tabular} { l  c c}
\hline
 Parameter &  Planck & Euclid+Planck\\
\hline
{\boldmath$10^{-2}\omega_\mathrm{b }$} 	&	 $2.210\pm 0.014            $	&	 $2.210\pm 0.011            $	\\
{\boldmath$\omega_\mathrm{cdm }$} 	&	 $0.11998\pm 0.00095        $	&	 $0.12005\pm 0.00027        $	\\
{\boldmath$100\theta_\mathrm{s }$} 	&	 $1.04111\pm 0.00033        $	&	 $1.04111\pm 0.00031        $	\\
{\boldmath$\tau_\mathrm{reio }$} 	&	 $0.0884^{+0.0045}_{-0.0054}$	&	 $0.0891\pm 0.0031          $	\\
{\boldmath$\ln(10^{10}V_0)$} 	&	 $1.7321^{+0.0089}_{-0.011} $	&	 $1.7331\pm 0.0058          $	\\
{\boldmath$\phi_{0}       $} 	&	 $< 0.0197                  $	&	 $< 0.0197                  $	\\
{\boldmath$\gamma         $} 	&	 Unbounded                         	&	 Unbounded                         	\\
{\boldmath$\phi_\mathrm{T}$} 	&	 $< 7.81                    $	&	 $< 7.80                    $	\\
{\boldmath$\ln\delta      $} 	&	 $-4.28^{+1.7}_{-0.93}      $	&	 $-4.40^{+1.6}_{-0.80}      $	\\
{\boldmath$\Omega_{\Lambda }$} 	&	 $0.6804\pm 0.0060          $	&	 $0.6801\pm 0.0013          $	\\
{\boldmath$\Omega_\mathrm{m }$} 	&	 $0.3195\pm 0.0060          $	&	 $0.3198\pm 0.0013          $	\\
{\boldmath$H_0            $} 	&	 $66.84\pm 0.44             $	&	 $66.82\pm 0.13             $	\\
{\boldmath$\sigma_8       $} 	&	 $0.8304\pm 0.0051          $	&	 $0.8313\pm 0.0018          $	\\
\hline
\end{tabular}

	\end{table}
	
\begin{figure*}
	\includegraphics[width=168mm]{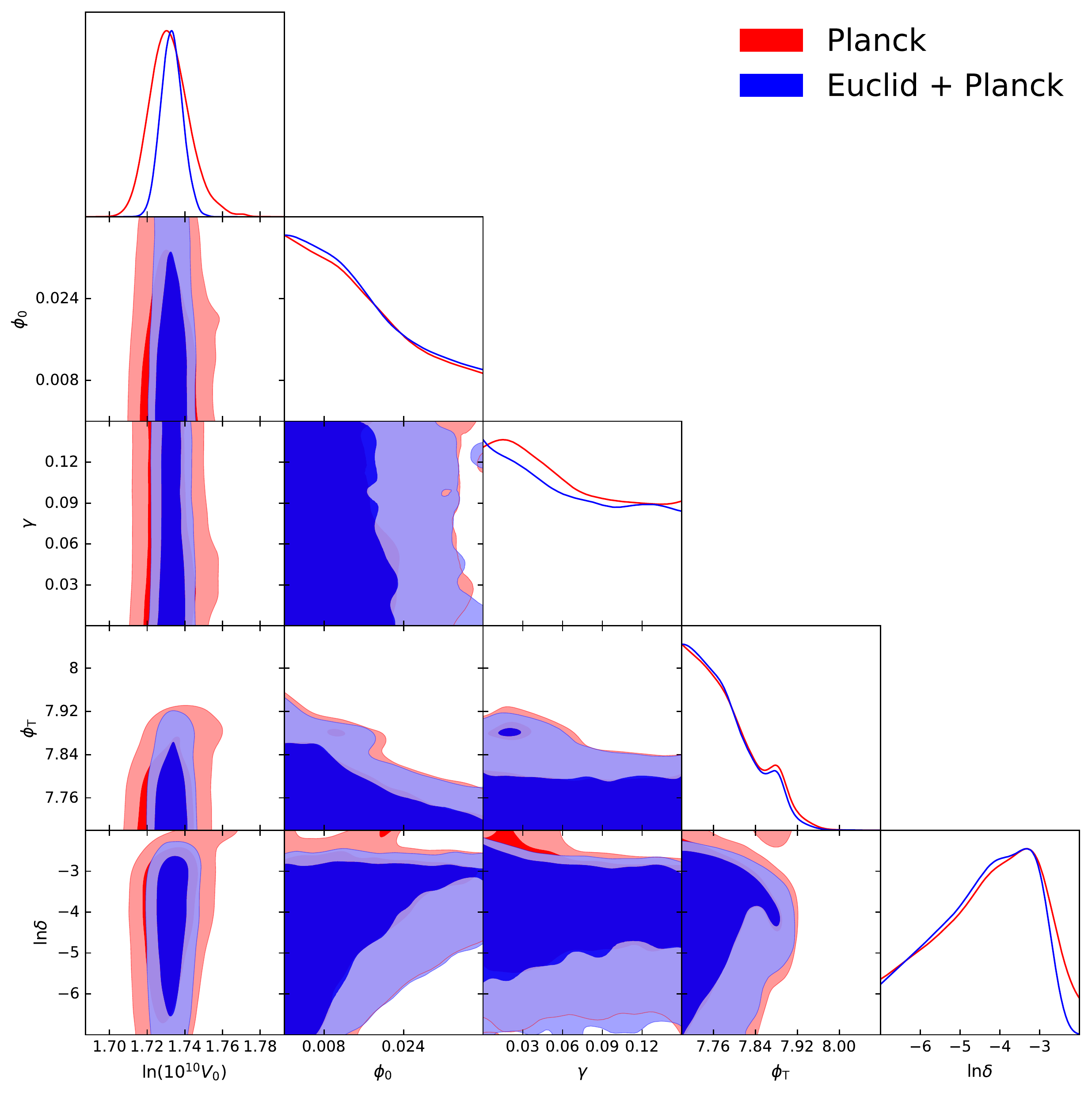}
    \caption{One-dimensional posteriors and marginalized contours for the inflation parameters in the WWI--A model.
     Improvement in the constraints is most evident in the amplitude parameter.}
    \label{fig:WWIA}
\end{figure*}

\begin{table}
	\centering
	\caption{$1 \sigma$ confidence intervals for cosmological parameters with WWI--B as the fiducial cosmology.}
	\label{tab:WWIB}
	
\begin{tabular} { l c c}
\hline
 Parameter &  Planck & Euclid+Planck\\
\hline
{\boldmath$10^{-2}\omega_\mathrm{b }$} 	&	 $2.209\pm 0.014            $	&	 $2.210\pm 0.011            $	\\
{\boldmath$\omega_\mathrm{cdm }$} 	&	 $0.1199\pm 0.0010          $	&	 $0.12000\pm 0.00025        $	\\
{\boldmath$100\theta_\mathrm{s }$} 	&	 $1.04110\pm 0.00032        $	&	 $1.04110\pm 0.00029        $	\\
{\boldmath$\tau_\mathrm{reio }$} 	&	 $0.0869^{+0.0045}_{-0.0064}$	&	 $0.0899\pm 0.0027          $	\\
{\boldmath$\ln(10^{10}V_0)$} 	&	 $1.7479^{+0.0086}_{-0.013} $	&	 $1.7536\pm 0.0049          $	\\
{\boldmath$\phi_{0}       $} 	&	 $< 0.0140                  $	&	 $0.00388^{+0.00040}_{-0.00051}$	\\
{\boldmath$\gamma         $} 	&	 Unbounded                        	&	 $0.041^{+0.019}_{-0.022}   $	\\
{\boldmath$\phi_\mathrm{T}$} 	&	 $< 7.84                    $	&	 $7.91056^{+0.00030}_{-0.00025}$	\\
{\boldmath$\ln\delta      $} 	&	 $> -6.19                   $	&	 $-7.05\pm 0.18             $	\\
{\boldmath$\Omega_{\Lambda }$} 	&	 $0.6807\pm 0.0063          $	&	 $0.68033\pm 0.00085        $	\\
{\boldmath$\Omega_\mathrm{m }$} 	&	 $0.3193\pm 0.0063          $	&	 $0.31958\pm 0.00085        $	\\
{\boldmath$H_0            $} 	&	 $66.86\pm 0.46             $	&	 $66.833\pm 0.091           $	\\
{\boldmath$\sigma_8       $} 	&	 $0.8370\pm 0.0052          $	&	 $0.8397\pm 0.0016          $	\\
\hline
\end{tabular}

	\end{table}

\begin{figure*}
	\includegraphics[width=168mm]{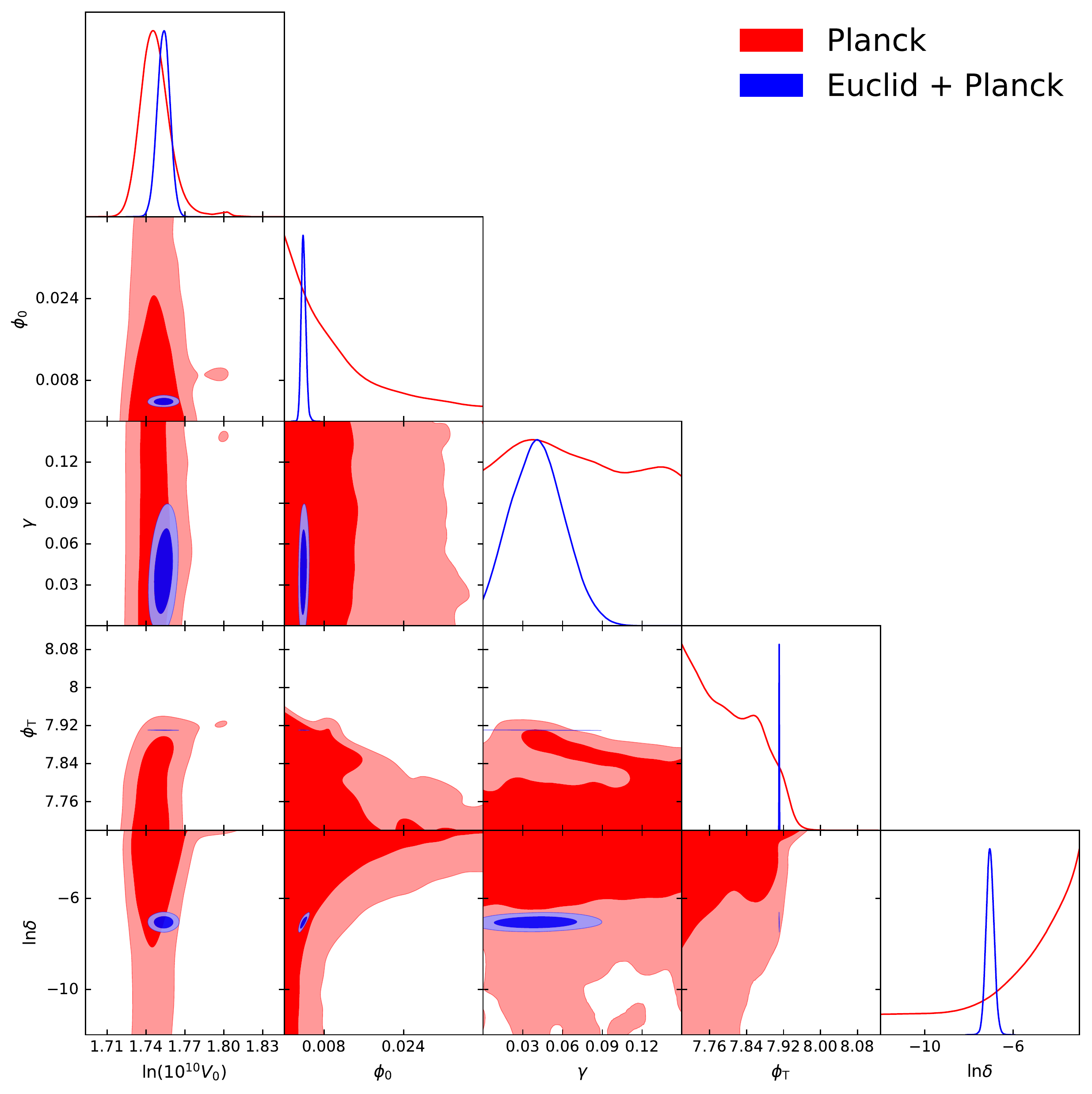}
    \caption{One-dimensional posteriors and marginalized contours for the inflation parameters in the WWI--B model. 
    The improvement in constraints for all inflation parameters with the addition of \textit{Euclid} data is evident in this plot.}
    \label{fig:WWIB}
\end{figure*}

\begin{table}
	\centering
	\caption{$1 \sigma$ confidence intervals for cosmological parameters with WWI--C as the fiducial cosmology. }
	\label{tab:WWIC}
	
\begin{tabular} { l c c}
\hline
 Parameter &  Planck & Euclid+Planck\\
\hline
{\boldmath$10^{-2}\omega_\mathrm{b }$} 	&	 $2.210\pm 0.014            $	&	 $2.211^{+0.011}_{-0.012}   $	\\
{\boldmath$\omega_\mathrm{cdm }$} 	&	 $0.12013\pm 0.00089        $	&	 $0.12004\pm 0.00027        $	\\
{\boldmath$100\theta_\mathrm{s }$} 	&	 $1.04110\pm 0.00032        $	&	 $1.04112\pm 0.00031        $	\\
{\boldmath$\tau_\mathrm{reio }$} 	&	 $0.0889^{+0.0046}_{-0.0056}$	&	 $0.0894\pm 0.0031          $	\\
{\boldmath$\ln(10^{10}V_0)$} 	&	 $1.7176^{+0.0091}_{-0.011} $	&	 $1.7186\pm 0.0059          $	\\
{\boldmath$\phi_{0}       $} 	&	 $< 0.00844                 $	&	 $< 0.00868                 $	\\
{\boldmath$\gamma         $} 	&	 Unbounded                         	&	 $< 0.0919                  $	\\
{\boldmath$\phi_\mathrm{T}$} 	&	 $< 7.82                    $	&	 $< 7.81                    $	\\
{\boldmath$\ln\delta      $} 	&	 $> -7.36                   $	&	 $> -7.00                   $	\\
{\boldmath$\Omega_{\Lambda }$} 	&	 $0.6795\pm 0.0057          $	&	 $0.6802\pm 0.0013          $	\\
{\boldmath$\Omega_\mathrm{m }$} 	&	 $0.3204\pm 0.0057          $	&	 $0.3197\pm 0.0013          $	\\
{\boldmath$H_0            $} 	&	 $66.79\pm 0.42             $	&	 $66.83\pm 0.13             $	\\
{\boldmath$\sigma_8       $} 	&	 $0.8249\pm 0.0051          $	&	 $0.8249\pm 0.0019          $	\\

\hline
\end{tabular}

	\end{table}

\begin{figure*}
	\includegraphics[width=168mm]{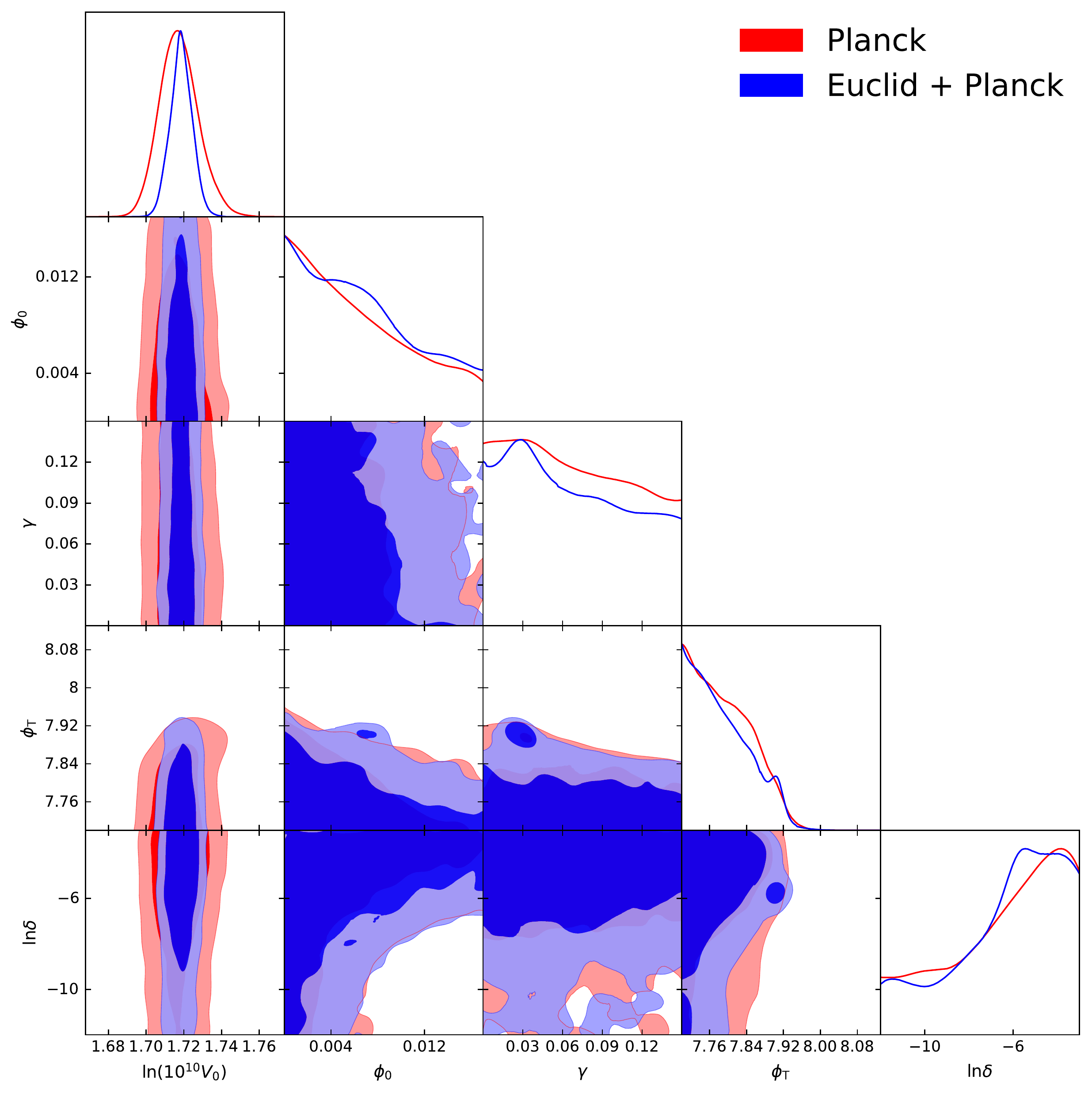}
    \caption{One-dimensional posteriors and marginalized contours for the inflation parameters in the WWI--C model. 
    \textit{Planck} alone provides no constraints on $\gamma$. The addition of \textit{Euclid} data results in an upper bound.}
    \label{fig:WWIC}
\end{figure*}

\begin{table}
	\centering
	\caption{$1 \sigma$ confidence intervals for cosmological parameters with WWI--D as the fiducial cosmology.}
	\label{tab:WWID}
	
\begin{tabular} { l  c c}
\hline
 Parameter &  Planck & Euclid+Planck\\
\hline
{\boldmath$10^{-2}\omega_\mathrm{b }$} 	&	 $2.204\pm 0.015            $	&	 $2.207\pm 0.011            $	\\
{\boldmath$\omega_\mathrm{cdm }$} 	&	 $0.12060\pm 0.00093        $	&	 $0.12037\pm 0.00028        $	\\
{\boldmath$100\theta_\mathrm{s }$} 	&	 $1.04093\pm 0.00032        $	&	 $1.04095\pm 0.00031        $	\\
{\boldmath$\tau_\mathrm{reio }$} 	&	 $0.0868^{+0.0056}_{-0.0069}$	&	 $0.0884\pm 0.0033          $	\\
{\boldmath$\ln(10^{10}V_0)$} 	&	 $1.751^{+0.011}_{-0.014}   $	&	 $1.7537\pm 0.0060          $	\\
{\boldmath$\phi_{0}       $} 	&	 $< 0.00756                 $	&	 $< 0.00815                 $	\\
{\boldmath$\gamma         $} 	&	 $0.071^{+0.034}_{-0.052}   $	&	 $0.072^{+0.034}_{-0.051}   $	\\
{\boldmath$\phi_\mathrm{T}$} 	&	 $< 7.89                    $	&	 $7.848^{+0.084}_{-0.14}    $	\\
{\boldmath$\ln\delta      $} 	&	 $> -6.66                   $	&	 $-5.49^{+2.4}_{-0.36}      $	\\
{\boldmath$\Omega_{\Lambda }$} 	&	 $0.6760\pm 0.0060          $	&	 $0.6776\pm 0.0014          $	\\
{\boldmath$\Omega_\mathrm{m }$} 	&	 $0.3239\pm 0.0060          $	&	 $0.3223\pm 0.0014          $	\\
{\boldmath$H_0            $} 	&	 $66.52\pm 0.43             $	&	 $66.63\pm 0.13             $	\\
{\boldmath$\sigma_8       $} 	&	 $0.8402\pm 0.0057          $	&	 $0.8406\pm 0.0019          $	\\

\hline
\end{tabular}

	\end{table}

\begin{figure*}
	\includegraphics[width=168mm]{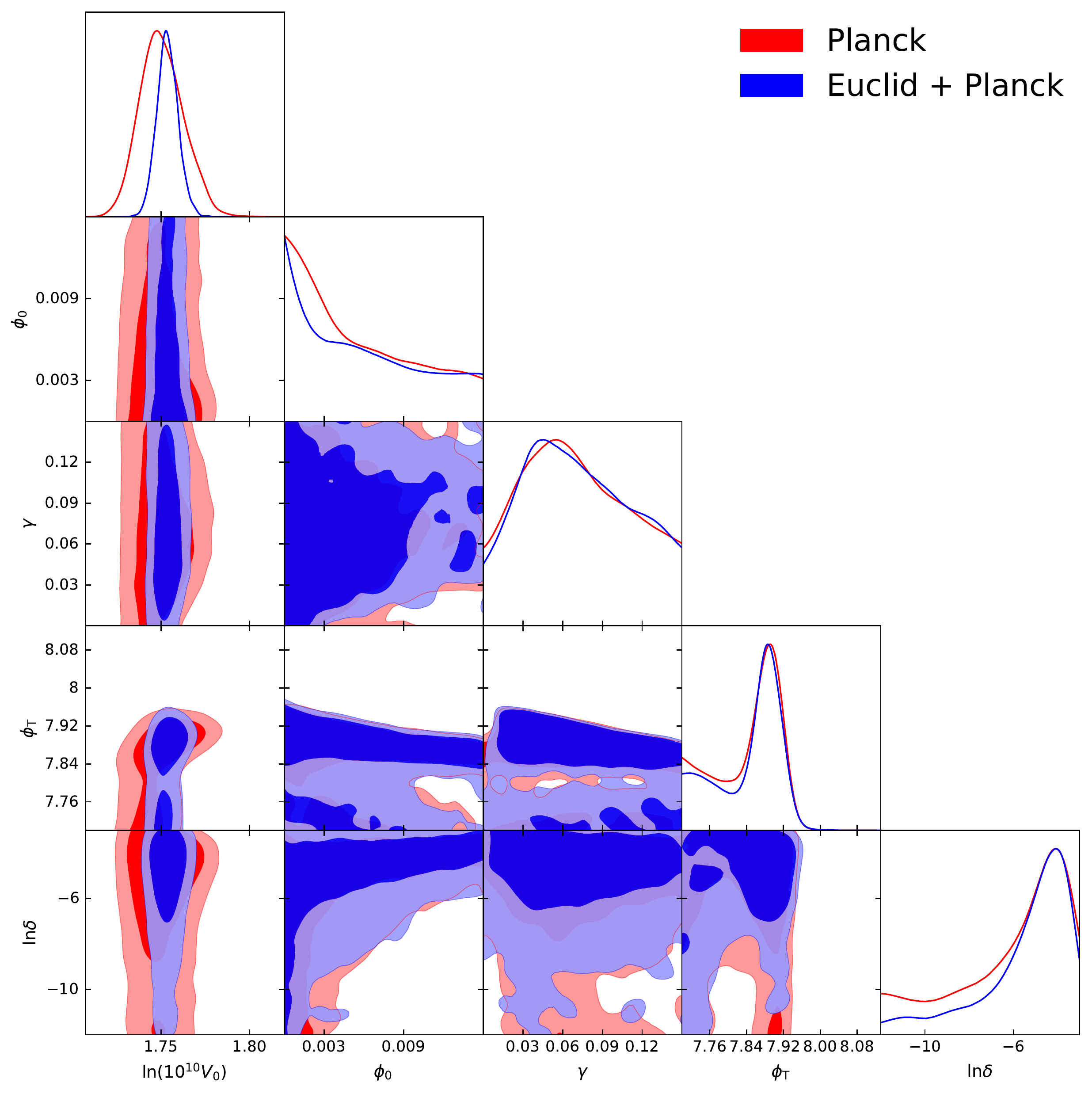}
    \caption{One-dimensional posteriors and marginalized contours for the inflation parameters in the WWI--D model.}
    \label{fig:WWID}
\end{figure*}

Compared to the 2-parameter from of power law primordial power spectrum, the WWI framework has 5 and 3 parameters that defines the potential of inflation. Since the feature induced by these parameters are scale-dependent, the power of \textit{Euclid} in constraining these parameters will be different in different fiducials. 
To begin with, we test the WWI-featureless fiducial obtained by keeping $\phi_0=\gamma=0$. This fiducial represents a nearly scale-invariant primordial spectrum with a spectral tilt of 0.96. When the WWI potential is compared against this fiducial using the combined \textit{Planck} + \textit{Euclid}  mock likelihood, we can address to what degree we can rule out non-zero $\phi_0,\gamma$, if a featureless power spectrum represents the true model. The 68 per cent constraints are provided in~\autoref{tab:WWIFeatureless} and constraints on inflation potential parameters are plotted in~\autoref{fig:WWIFeatureless}. While the constraints on the background parameters experience similar improvements with \textit{Euclid}  as the power law $\LCDM$ model, we find that apart from $V_0$, we do not have any improvements with respect to \textit{Planck}. In other words, if the primordial power spectrum does indeed follow power law, \textit{Euclid}  is not going to be able to rule out any large-scale power suppression (induced by $\gamma$) or oscillations (induced by $\phi_0$) with higher statistical significance than \textit{Planck} has already done. It is expected as \textit{Planck} being cosmic variance limited in temperature, provides the best constraints at the large scales ($\ell<50,~k<5\times10^{-3} \mathrm{Mpc}^{-1}$). At the same time, apart from high frequency oscillations \textit{Planck} already rules out wiggles at small scales. Therefore a featureless fiducial representing the mock data is not expected to rule out potentials that are already ruled out by Planck. For WWI-B and WWI-D we will consider high frequency wiggles in the primordial spectra as fiducials.  
With the feature models, we add 1 (for WWIP) or 4 (for WWI) free parameters to our MCMC parameter space with respect to the Concordance Model, which should induce weaker parameter constraints in this kind of fiducial-based forecast. The reason is simple: more free parameters means greater statistical uncertainty. However, we note that there is no significant degradation in the constraints on the background parameters in the feature models. 
Indeed we note an improvement in constraints on the CDM density compared to Concordance Model in both \textit{Planck} and Planck+Euclid. In the potential when we fix $\mu$ in~\autoref{eq:equation-WWI}, the spectral tilt generated by the inflation gets fixed. Therefore in the MCMC runs, variation in the tilt is not allowed and that reduces the degeneracies with the background parameter resulting in marginally improved constraints.

In \autoref{tab:WWIA} we present the constraints on the WWI potential when WWI-A is used as fiducial cosmology and in \autoref{fig:WWIA} we plot the constraints only on the inflation potential parameters. The table reflects improvement in constraints on the background parameters with \textit{Euclid} similar to the Concordance Model. However, we do not find any improvement in the potential parameters except $V_0$ which represents the amplitude of the primordial spectrum. Compared to the power law, WWI-A specifically improves the fit to the \textit{Planck} data at low multipoles ($\ell<20$) with the large-scale suppression, and the dip at near $k\sim2\times10^{-3}~\mathrm{Mpc}^{-1}$ fits the $\ell\sim22$ dip in the angular power spectrum. The power spectrum at these largest scales probed by \textit{Planck} can not be constrained better with \textit{Euclid}  data as these scales are dominated by cosmic variance, and we can only expect an improvement with cosmic-variance-limited polarization surveys~\cite{Hazra:coreforecast}. The \textit{Euclid} measurement error at the largest scales is dominated by statistical uncertainties due to cosmic variance. Cosmic-variance errors on the dark energy
equation of state \citep{Valkenburg2013} and the Hubble parameter \citep{Marra2013} are particularly important. Here we assume a cosmological constant, so we are not concerned by the former. But the latter effect may degrade the constraints on our other parameter through their degeneracies with $H_0$. These limitations can be reduced by using multiple tracers with different biases. We should note that both \textit{Planck} and \textit{Planck}+\textit{Euclid} reject the high amplitude (high $\phi_0$) sharp oscillations (low $\ln\delta$) as we can note from their correlations. A transition at higher field value ($\phi_\mathrm{T}$) implies the occurrence of features at small scales as the small-scale modes leave the Hubble radius at a higher $\phi_\mathrm{T}$ values. We have strong constraints in the increasing direction of both $\phi_0-\phi_T$ and $\gamma-\phi_T$. Therefore, we also find high amplitude oscillations and suppression are only allowed at large scales and up to certain intermediate scales. \textit{Euclid}  can marginally tighten the constraint on the frequency of the oscillation by constraining $\ln\delta$.

The WWI-B, C and D fiducial models represent wiggles in the primordial spectrum within intermediate to smaller scales ($\sim0.1\mathrm{Mpc}^{-1}$) as in the inset of~\autoref{fig:pk}. Note that these spectra fall within the high signal-to-noise region of both \textit{Planck} and \textit{Euclid}. Constraints on WWI model when WWI-B, WWI-C and WWI-D are used as fiducials are tabulated in~\autoref{tab:WWIB},~\autoref{tab:WWIC},~\autoref{tab:WWID}, respectively, and corresponding posteriors and marginalized contours are plotted in~\autoref{fig:WWIB},~\autoref{fig:WWIC},~\autoref{fig:WWID}. Out of these three cases we note a remarkable improvement in constraints for the WWI-B case when \textit{Euclid} is combined with \textit{Planck}. We obtain significant detection of $\phi_0$ and 1-2$\sigma$ preference of a suppression ($\gamma$). Therefore, if WWI-B represents the true model of our Universe, \textit{Euclid}  will certainly able to establish this with high statistical significance when \textit{Planck} CMB data is used in combination. Since out of the four WWI fiducials, WWI-B has the maximum large-scale suppression, using this fiducial leads to a marginal preference for $\gamma$. However, due to cosmic variance, it is not possible to get more than 2$\sigma$ preference with power spectrum. A detection of $\phi_0$ represents a detection of the wiggles in the primordial power spectrum and therefore we obtain the position of the potential transition ($\phi_\mathrm{T}$) and the sharpness of the transition ($\ln\delta$) determined with high statistical significance as well. WWI-C has wiggles in the intermediate scales but these oscillations decay at the smaller scales ($k\sim10^{-2}\mathrm{Mpc}^{-1}$). This limited overlap with \textit{Euclid}-probed cosmological scales reduces the chances of a detection of these features. When the WWI-D fiducial is used as mock data, \textit{Euclid}  improves the constraints on the inflationary parameters compared to \textit{Planck}-only results. We find that the constraints on the location and sharpness of the feature can experience a slight improvement. However, unlike WWI-B we will cannot expect any detection of features. In this case, although WWI-D has the best overlap between \textit{Planck} and \textit{Euclid}-probed cosmological scales, the amplitude of the oscillations is the lowest among the WWI models and therefore is less likely to be detected by \textit{Euclid} . Note that with CORE-like surveys it was found that the WWI-D spectrum has the best chance of being detected~\citet{Hazra:coreforecast}. However, in this case \textit{Euclid} is not expected to resolve the high frequency oscillations as they will be binned and averaged out in the observed power spectrum.

\begin{table}
	\centering
	\caption{$1 \sigma$ confidence intervals for the inflation parameters in the WWIP:Featureless model. }
	\label{tab:WWIPfeatureless}
	
\begin{tabular} { l c  c}
\hline
 Parameter &  Planck & Euclid+Planck\\
\hline
{\boldmath$10^{-2}\omega_\mathrm{b }$} 	&	 $2.211\pm 0.014            $	&	 $2.212\pm 0.011            $\\
{\boldmath$\omega_\mathrm{cdm }$} 	&	 $0.12012\pm 0.00089        $	&	 $0.11992\pm 0.00027        $\\
{\boldmath$100\theta_\mathrm{s }$} 	&	 $1.04115\pm 0.00033        $	&	 $1.04116\pm 0.00031        $\\
{\boldmath$\tau_\mathrm{reio }$} 	&	 $0.0918^{+0.0045}_{-0.0051}$	&	 $0.0910^{+0.0028}_{-0.0032}$\\
{\boldmath$\ln(10^{10}V_0)$} 	&	 $0.2857^{+0.0090}_{-0.010} $	&	 $0.2839\pm 0.0057          $\\
{\boldmath$\phi_{0}       $} 	&	 $< 0.450                   $	&	 $< 0.416                   $\\
{\boldmath$\phi_\mathrm{T}$} 	&	 $4.590^{+0.044}_{-0.032}   $	&	 $> 4.57                    $\\
{\boldmath$\Omega_{\Lambda }$} 	&	 $0.6798\pm 0.0057          $	&	 $0.6811\pm 0.0013          $\\
{\boldmath$\Omega_{m }    $} 	&	 $0.3201\pm 0.0057          $	&	 $0.3188\pm 0.0013          $\\
{\boldmath$H_0            $} 	&	 $66.81\pm 0.42             $	&	 $66.90\pm 0.13             $\\
{\boldmath$\sigma_8       $} 	&	 $0.8352\pm 0.0049          $	&	 $0.8337\pm 0.0018          $\\
\hline
\end{tabular}

	\end{table}

\begin{figure*}
	\includegraphics[width=168mm]{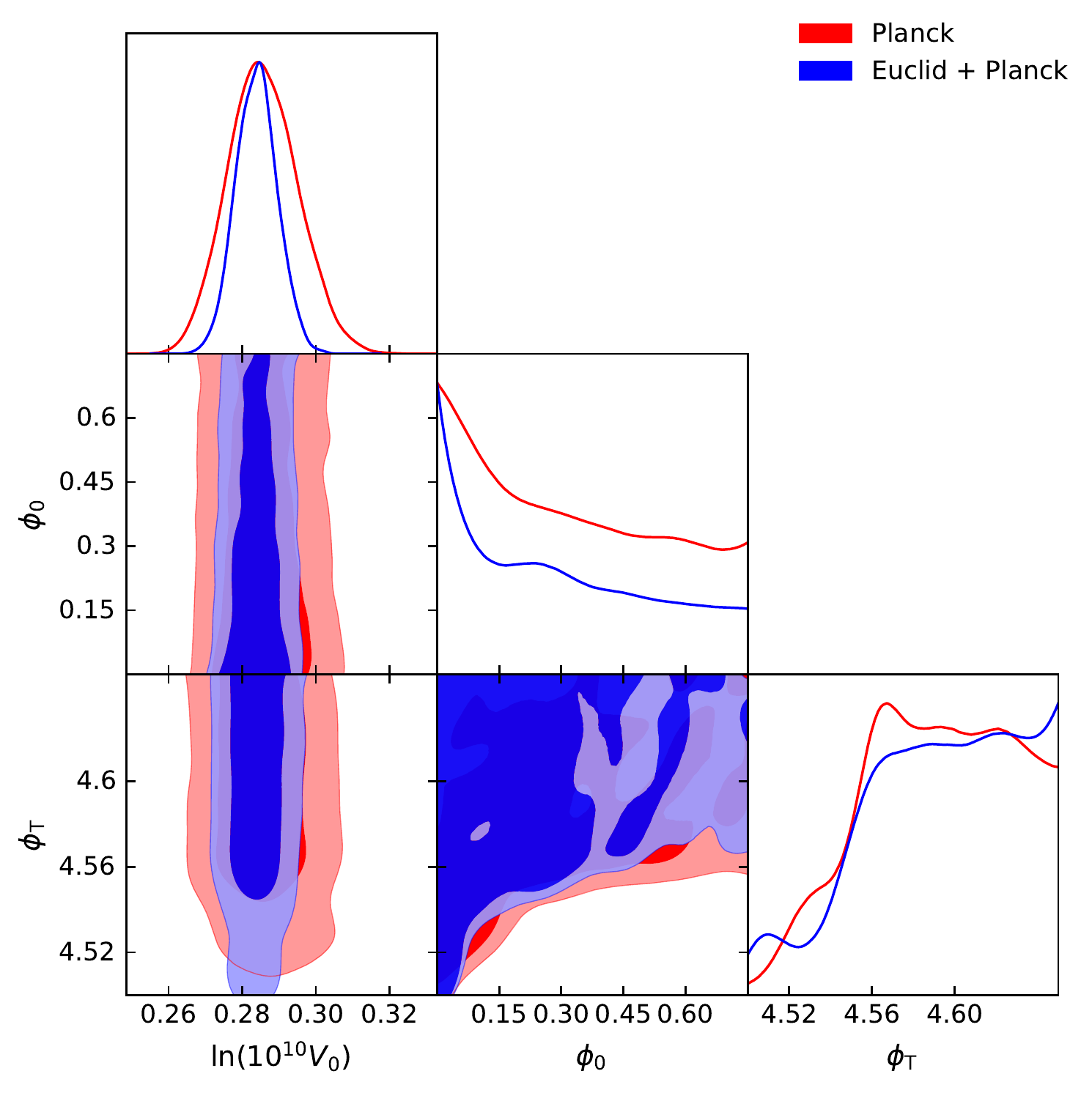}
	    \caption{One-dimensional posteriors and marginalized contours for the inflation parameters in the WWIP:Featureless model.}
    \label{fig:WWIP1}
\end{figure*}

\begin{table}
	\centering
	\caption {$1 \sigma$ confidence intervals for cosmological parameters with WWIP:Planck-best-fit as the fiducial cosmology.}
	\label{tab:WWIP2}
	
\begin{tabular} { l c c}
\hline
 Parameter &  Planck & Euclid+Planck\\
\hline
{\boldmath$10^{-2}\omega_\mathrm{b }$} 	&	 $2.209\pm 0.014            $	&	 $2.209\pm 0.012            $\\
{\boldmath$\omega_\mathrm{cdm }$} 	&	 $0.12010\pm 0.00089        $	&	 $0.12009^{+0.00030}_{-0.00024}$\\
{\boldmath$100\theta_{s } $} 	&	 $1.04109\pm 0.00032        $	&	 $1.04112\pm 0.00030        $\\
{\boldmath$\tau_\mathrm{reio }$} 	&	 $0.0860^{+0.0044}_{-0.0054}$	&	 $0.0883\pm 0.0030          $\\
{\boldmath$\ln(10^{10}V_0)$} 	&	 $0.2743^{+0.0087}_{-0.011} $	&	 $0.2790\pm 0.0056          $\\
{\boldmath$\phi_{0}       $} 	&	 $< 0.435                   $	&	 $< 0.424                   $\\
{\boldmath$\phi_\mathrm{T}$} 	&	 $> 4.57                    $	&	 $> 4.57                    $\\
{\boldmath$\Omega_{\Lambda }$} 	&	 $0.6797\pm 0.0057          $	&	 $0.6798^{+0.0013}_{-0.0012}$\\
{\boldmath$\Omega_\mathrm{m }$} 	&	 $0.3203\pm 0.0057          $	&	 $0.3201^{+0.0012}_{-0.0013}$\\
{\boldmath$H_0            $} 	&	 $66.79\pm 0.41             $	&	 $66.80\pm 0.12             $\\
{\boldmath$\sigma_8       $} 	&	 $0.8307\pm 0.0050          $	&	 $0.8327\pm 0.0018          $\\

\hline
\end{tabular}

	\end{table}

\begin{figure*}
	\includegraphics[width=168mm]{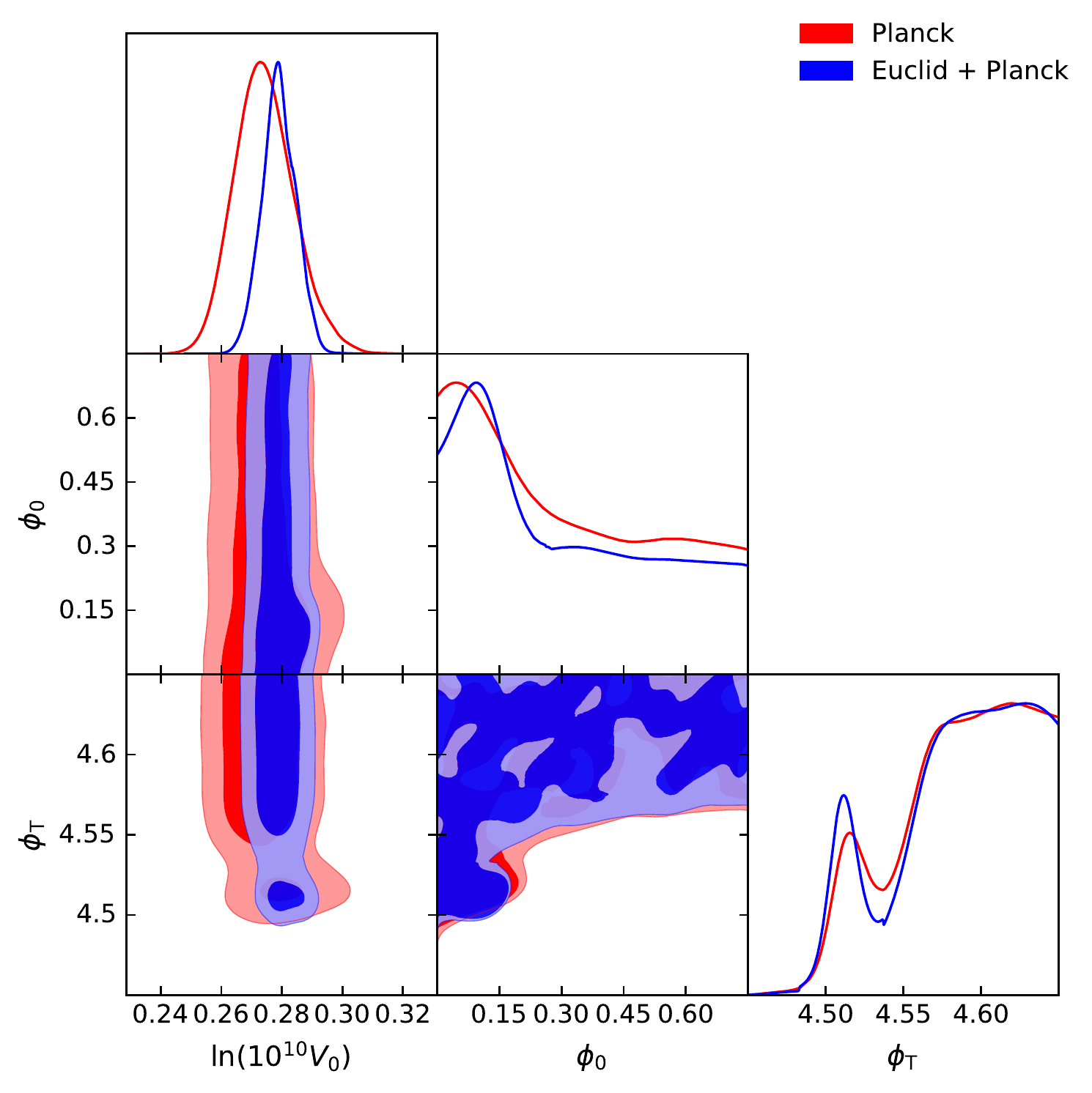}
    \caption{One-dimensional posteriors and marginalized contours for the inflation parameters in the WWIP:Planck-best-fit model.}
    \label{fig:WWIP2}
\end{figure*}

\begin{table}
	\centering
	\caption{$1 \sigma$ confidence intervals for cosmological parameters with WWIP:Small-scale-feature as the fiducial cosmology.  }
	\label{tab:WWIP3}
	\begin{tabular} { l  c c}
\hline
 Parameter &  Planck & Planck+Euclid\\
\hline
{\boldmath$10^{-2}\omega_\mathrm{b }$} 	&	 $2.210\pm 0.014            $	&	 $2.210^{+0.010}_{-0.012}   $\\
{\boldmath$\omega_\mathrm{cdm }$} 	&	 $0.11999\pm 0.00090        $	&	 $0.12001\pm 0.00026        $\\
{\boldmath$100\theta_\mathrm{s }$} 	&	 $1.04110\pm 0.00033        $	&	 $1.04111\pm 0.00030        $\\
{\boldmath$\tau_\mathrm{reio }$} 	&	 $0.0903\pm 0.0054          $	&	 $0.0899\pm 0.0029          $\\
{\boldmath$\ln(10^{10}V_0)$} 	&	 $0.301\pm 0.011            $	&	 $0.3000\pm 0.0054          $\\
{\boldmath$\phi_{0}       $} 	&	 $0.177\pm 0.037            $	&	 $0.178\pm 0.023            $\\
{\boldmath$\phi_\mathrm{T}$} 	&	 $4.50246^{+0.00043}_{-0.00039}$	&	 $4.50245\pm 0.00017        $\\
{\boldmath$\Omega_{\Lambda }$} 	&	 $0.6804\pm 0.0057          $	&	 $0.6803\pm 0.0012          $\\
{\boldmath$\Omega_\mathrm{m }$} 	&	 $0.3196\pm 0.0057          $	&	 $0.3196\pm 0.0012          $\\
{\boldmath$H_0            $} 	&	 $66.84\pm 0.42             $	&	 $66.84\pm 0.12             $\\
{\boldmath$\sigma_8       $} 	&	 $0.8413\pm 0.0053          $	&	 $0.8408\pm 0.0018          $\\

\hline
\end{tabular}

	\end{table}

\begin{figure*}
	\includegraphics[width=168mm]{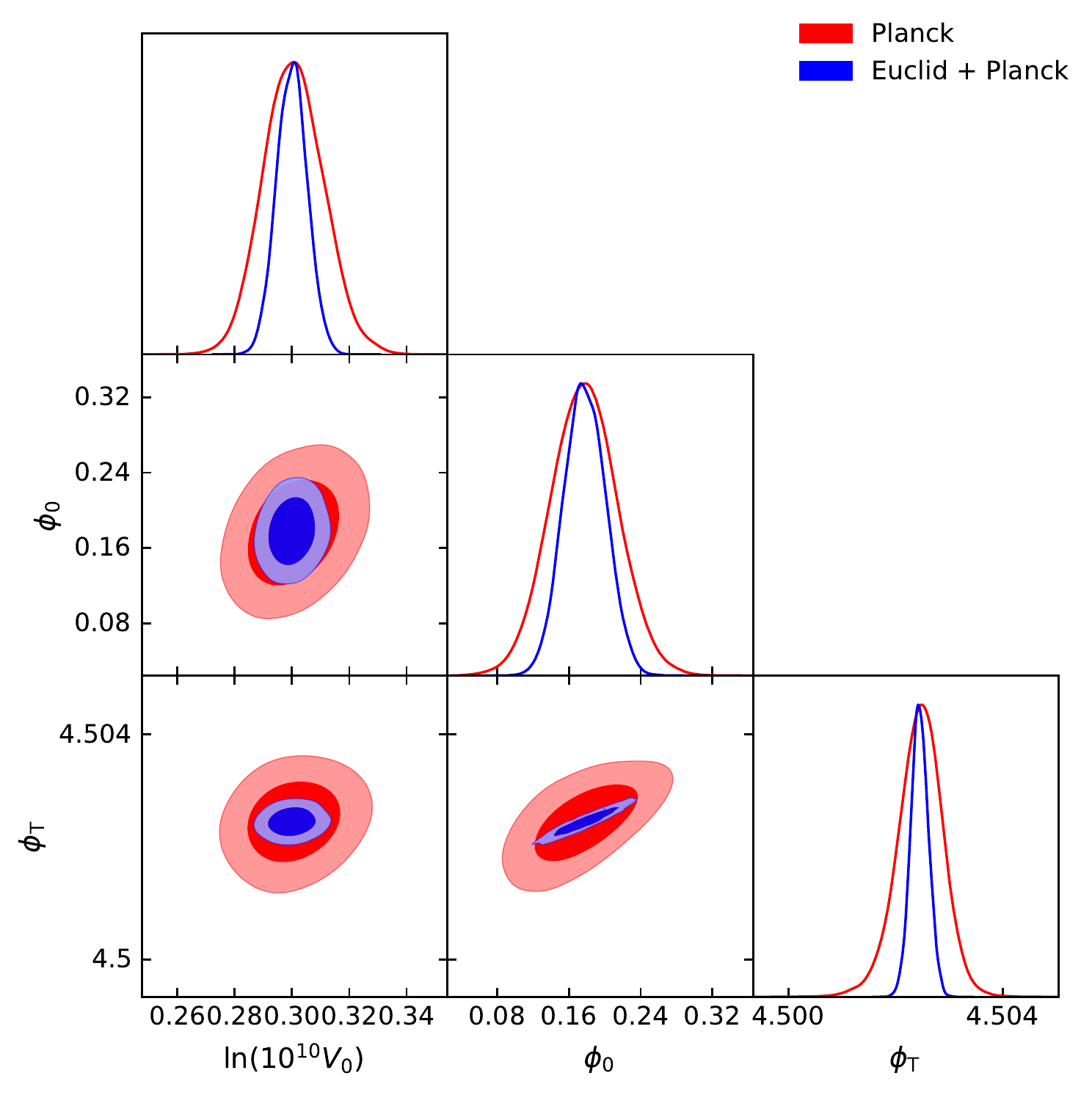}
    \caption{One-dimensional posteriors and marginalized contours for the inflation parameters in the WWIP:Small-scale-feature model. We obtain closed contours for all the inflation parameters, with a significant improvement with \textit{Euclid} data are added.}
    \label{fig:WWIP3}
\end{figure*}

The WWIP potential has 3 parameters describing the primordial physics. Apart from the amplitude, determined by $V_0$, two other parameters $\phi_0$ and $\phi_\mathrm{T}$ are responsible for the transition in the potential and therefore for features. Note that this model produces both suppression and wiggles. We have used three fiducial primordial spectra to generate the mock spectra for \textit{Euclid}  and Planck. Similar to WWI potential, we use a featureless fiducial generated with $\phi_0=0$. For the second fiducial we use the best fit to \textit{Planck} temperature and polarization data~\citep{Hazra2016}. Another point in the potential parameter space that is allowed within \textit{Planck} 95 per cent confidence limits and generates features extended towards smaller scales ($k\sim0.2$ in $h\mathrm{Mpc}^{-1}$) in the primordial spectrum (compared to WWIP:Planck-best-fit), is used here as mock data termed as WWIP:small-scale-feature. The constraints on the background and inflationary parameters for these three cases are presented in~\autoref{tab:WWIPfeatureless},~\autoref{tab:WWIP2},~\autoref{tab:WWIP3}, respectively. The one-dimensional posteriors and marginalized contours for the inflationary potential parameters are plotted in~\autoref{fig:WWIP1},~\autoref{fig:WWIP2} and~\autoref{fig:WWIP3} respectively. Similar to the results obtained so far, we notice 40\% improvement in the constraints on $V_0$ (in logscale). In the featureless case we notice 10\% improvement by \textit{Euclid} with respect to \textit{Planck} on the upper bound on $\phi_0$ (the amplitude of the wiggles). 
In \autoref{tab:WWIPfeatureless}, we note that adding \textit{Euclid} information shifts the mean value of $\phi_\mathrm{T}$ closer to the upper edge of the prior boundary, so the binning method for the analysis of the MCMC chains is unable to find two-tailed marginalized limits, hence the single (lower-only) bound for $\phi_\mathrm{T}$ obtained with \textit{Euclid}+\textit{Planck}.  
In the case where the WWIP:Planck-best-fit represents the true model of the Universe, we find only marginal improvement in $\phi_0$ as can be seen in~\autoref{fig:WWIP2}. The current \textit{Planck} best fit for WWIP has oscillations in the large to intermediate scales ($k\sim10^{-3} - 10^{-2}\mathrm{Mpc}^{-1}$). This range of scales is already well-probed by \textit{Planck} and from \textit{Euclid}  we only expect to see marginal improvement. However, if WWIP:small-scale-feature represents a true model of the Universe, we can expect 40 per cent improvement in the constraints on $\phi_0$ leading to a detection of features with \textit{Euclid} +Planck (note that with \textit{Planck} mock also we are able to rule out featureless spectrum with more than $4\sigma$ confidence). Since WWIP:small-scale-feature has oscillations with higher magnitude and they extend to smaller scales with better overlap with \textit{Euclid}-probed scales, this improvement is expected.

Analysis with these fiducials clearly establishes the contribution of \textit{Euclid} data in constraining inflationary parameters, but the improvement in constraints varies. Projected constraints depend on the model and on the fiducial cosmology, due to the non-Gaussian nature of the posteriors. Since inflation features appear at particular scales, the overlap of these features with scale probed by \textit{Euclid} determines the improvements in constraints with respect to \textit{Planck} CMB data. In the cases where wiggles in the primordial spectrum are located in the intermediate to small scales ($k\sim10^{-3} - 10^{-1}\mathrm{Mpc}^{-1}$), \textit{Euclid}  can play a significant role in detection along with \textit{Planck} CMB. 

A consistent feature of the results for all models is the improvement in constraints on the amplitude of the matter power spectrum ($A_\mathrm{s}$ in the Concordance Model, or through $V_0$ in WWI). Euclid spectroscopy provides better redshift resolution, which results in a better measurement of the redshift-space distortion signal. It therefore breaks the degeneracy between the bias parameter and the amplitude of the power spectrum, leading to better constraints on the amplitude.

\section{Conclusions}
\label{sec:Conclusions}

We present accurate and realistic forecasts for \textit{Euclid} cosmic shear and galaxy clustering based on MCMC simulations for features in the primordial power spectrum. We use a $\LCDM$ background cosmology including massive neutrinos. The features in the primordial power spectra that we consider are both local and non-local in nature and we use the WWI framework, which can produce different kinds of features relevant to the CMB data observed by \textit{Planck}. With a discontinuity either in the potential or in its derivative, WWI provides large-scale suppression, 
localized and non-local oscillations. Our results compliment those in \citet{2012JCAP...04..005H} and \citet{2016JCAP...10..041B}, where an MCMC-based forecast was carried out for galaxy clustering data from different probes including \textit{Euclid}. \citet{2018JCAP_Ballardini} report similar results based on Fisher analysis.
We use up-to-date specifications for the \textit{Euclid} survey, and recently-published likelihoods for cosmic shear and galaxy clustering, with a conservative model for the theoretical error in the non-linear spectrum, with a redshift-dependent cut-off at $0.2\, h\text{Mpc}^{-1}$. By using this realistic error model with MCMC simulations, we are free from the assumption of Gaussianity which is the basis of Fisher analysis. For this reason, Fisher analysis tends to underestimate the error bounds.

Our results are in broad agreement with other studies such as \citep{Audren2013,Sprenger2018}. We show that the addition of \textit{Euclid} data tightens the cosmological parameter constraints obtained by \textit{Planck} alone, even with a `conservative' setup for the non-linear uncertainties. The results strengthen the scientific case for \textit{Euclid} and the use of multiple probes to exploit synergies and break parameter degeneracies.

For features at the very largest scales, the contribution by large-scale structure to the constraints is minimal, and most of the constraining power comes from CMB data. The improvement with the addition of \textit{Euclid} data mainly occurs for models with features at intermediate and small scales. For a more realistic cutoff, possibly with a decrease in the theoretical error through a more accurate prescription for the non-linear correction, we expect the addition of information from smaller scales to result in a tightening of cosmological constraints, although we have not quantified this improvement. For a power-law model, \citet{Sprenger2018} show that galaxy clustering is slightly more sensitive to the tilt of the primordial power spectrum, while cosmic shear is more sensitive to the amplitude.
 
Our conservative forecast establishes that MCMC is necessary tool in the detection of features in the primordial power spectrum, as well as in obtaining accurate forecasts. Simulated data is a single realisation of the {\it true} Universe, and we find a significant change in the inflation parameter constraints depending on the amplitude and occurrence of features at different cosmological scales. This hints at a strong parameter dependence of the joint \textit{Planck}-\textit{Euclid} covariance matrix in the inflation sector, which means that the Fisher matrix approach may not be sufficiently accurate. This parameter dependence merits further study (see e.g. \citealt{2017MNRAS.472.4244H}; \citealt{2019OJAp_Kodwani}; and, for the \textit{Euclid} dark energy Figure-of-Merit, \citealt{Debono2014};  \citealt{2016MNRAS.460.3398S}).

Our main findings are: 
\begin{enumerate}
\item The \textit{Euclid} cosmic shear and galaxy clustering likelihoods and error modelling by \citep{Sprenger2018} perform well with cosmological models containing features in the primordial power spectrum.
\item We find significant improvement in the constraints on the background parameters when \textit{Euclid} is used with \textit{Planck} compared to \textit{Planck} alone. The improvement for $\Omega_\mathrm{b}h^2$ is marginal, since it is already well-constrained by Planck.
\item Using cosmic shear with galaxy clustering, \textit{Euclid} is expected to improve the bounds on the primordial spectrum amplitude and tilt by 30 to 40 per cent compared to \textit{Planck} when the power-law form is used for the spectrum. 
Due to these tighter constraints, we also find indirect tighter bounds on the reionization optical depth. 
\item When WWI models of inflation are used for the forecast, the scale of the potential, $V_0$ gets tightly constrained in all the cases, as the amplitude of the perturbation spectrum is directly dependent on $V_0$. The constraints on the background cosmology parameters are not significantly affected by the presence of features in the primordial power spectrum.  
\item Features that are present at larger scales compared to our conservative large scale cutoff for Euclid ($0.02~\mathrm{Mpc}^{-1}$), cannot be constrained better than \textit{Planck} with `conservative' theoretical errors for \textit{Euclid}. However, we notice marginal improvement in some cases, where better constraints on background and amplitude parameters help to reduce certain residual degeneracies.
\item Oscillations that are present at intermediate and small scales ($k\sim0.02-0.2~\mathrm{Mpc}^{-1}$) in the power spectrum with $\sim 2$ per cent amplitude with respect to the featureless spectrum, have a high probability of being detected with high statistical significance with combined \textit{Euclid} and \textit{Planck} data, if they represent the true model of the Universe.
\item The contribution of \textit{Euclid} data to the detection of small-scale, high-frequency features is limited.
\end{enumerate}

Our work validates the scientific potential of \textit{Euclid} by contributing three main results.
\begin{enumerate}
\item First, our forecasts show that \textit{Euclid} improves constraints in the background cosmology sector, even in the presence of features in the inflation potential.
\item Second, we show that \textit{Euclid} data improves constraints in the overall scale of the slow-roll potential. This widens the scientific scope of \textit{Euclid} beyond the original dark energy and neutrino sectors. 
\item Third, we provide the first \textit{Euclid} forecasts using MCMC for the fine shape of the inflation potential in the presence of features. As the next generation of probes are activated, and the synergies between probes exploited, this is likely to become an important area of research. These include complementarity with the LSST, which could improve the signal, especially from faint galaxies \citep{Rhodes:2017,Robertson:2019},
 and the SKA and CMB-S4, which could provide independent validation of systematics as well as extra information from cross-correlation \citep{2018JCAP_Ballardini,2019BAAS, Sprenger2018}. 
 \textit{Planck} pushed the boundaries of CMB observations to the point where future experiments cannot do better by using the same data, so new experiments must turn to new observables (CMB polarization, CMB lensing, and other secondary effects). \textit{Euclid}'s ability to probe the primordial Universe though large-scale structure is a major milestone in observational cosmology. This work underlines the need for multiple probes in order to explore all cosmological scales as fully as possible. 
\end{enumerate}

As \textit{Euclid} likelihood modelling is improved closer to the launch date, it will be straightforward to obtain more accurate forecasts using our pipeline. Further improvements include, but are not limited to: better modelling of the non-linear theoretical error, especially in the presence of features in the primordial power spectrum; more accurate galaxy number counts; improved modelling of massive neutrino effects. The availability of simulated data from Wiggly Whipped Model would open up new avenues of research. Additional constraining power can be provided by independent probes such as SKA 21-cm intensity mapping from reionization, from priors on the Hubble parameter through supernovae, and also from cross-correlations between galaxy clustering and cosmic shear. The calculation of power spectra from the inflaton potential provides new opportunities for testing competing models,
and for Bayesian model selection with \textit{Euclid}. 

\section*{Acknowledgements}

Ivan Debono acknowledges that the research work disclosed in this publication is partially funded by the 
REACH HIGH Scholars Programme -- Post-Doctoral Grants. The grant is part-financed by the European Union, Operational Programme II --
Cohesion Policy 2014--2020. 
Dhiraj Kumar Hazra has received funding from the European Union’s Horizon 2020 research and innovation programme under the Marie Sklodowska-Curie grant agreement  No. 664931.
Alexei A. Starobinsky was partly supported by the project number 0033-2019-0005 of the
Russian Ministry of Science and Higher Education.
Ivan Debono would like to thank Thejs Brinckmann and Maria Archidiacono for useful discussions.
The authors gratefully acknowledge support from the CNRS/IN2P3 Computing Centre (Lyon -- France) for providing computing resources needed for this work.

\section*{Data availability}
The data underlying this article will be shared on reasonable request to the corresponding author.

\bibliographystyle{mnras}
\bibliography{Euclidinflation_Debono_Hazra_etal}

\label{lastpage}

\end{document}